\documentclass[brazilian,twocolumn,aps,prb,superscriptaddress,amsmath,amssymb,showpacs,floatfix,preprintnumbers,longbibliography]{revtex4-2}
\usepackage[latin9,utf8]{inputenc}
\usepackage{float}
\usepackage{textcomp}
\usepackage{amsmath,amssymb,amscd,hyperref}
\usepackage{graphicx}
\makeatletter
\usepackage{url}
\usepackage{grffile}
\usepackage{bbold}
\usepackage{comment}
\usepackage{dirtytalk}
\usepackage{chngcntr}
\usepackage{tabularx}
\usepackage{multirow}
\usepackage{soul}
\usepackage{cleveref}
\usepackage{braket}
\DeclareUnicodeCharacter{2212}{-}
\DeclareUnicodeCharacter{0301}{\hspace{-1ex}\'{ }}

\usepackage{dcolumn}
\usepackage{color}
\DeclareGraphicsExtensions{.png .jpg .pdf}
\usepackage{hyperref}
\hypersetup{
     colorlinks = true,
     linkcolor = blue,
     anchorcolor = blue,
     citecolor = blue,
     filecolor = blue,
     urlcolor = blue
     }
\usepackage{braket}
\usepackage{physics}
\usepackage{array}
\usepackage{booktabs}
\usepackage{soul}
\makeatother

\begin{document}

\title{Tilted chiral spin textures in confined nanostructures with in-plane magnetic anisotropy}

\author{W. L. Fu}
\affiliation{School of Materials and Physics, China University of Mining and Technology, Xuzhou 221116, P. R. China}
\affiliation{School of Integrated Circuits and Electronics, MIIT Key Laboratory for Low-Dimensional Quantum Structure and Devices, Beijing Institute of Technology, Beijing 100081, China}

\author{H. M. Dong}
\email{hmdong@cumt.edu.cn}
\affiliation{School of Materials and Physics, China University of Mining and Technology, Xuzhou 221116, P. R. China}

\author{K. Chang}
\email{kchang@zju.edu.cn}
\affiliation{School of Physics, Zhejiang University, Hangzhou 310027, P. R. China}

\date{\today}

\begin{abstract}
We demonstrate that nanoconfinement effects and in-plane magnetic anisotropy (IMA) can lead to tilted chiral spin textures in magnetic nanostructures, based on the analysis and simulation of theoretical models of micromagnetism. The tilted skyrmions are induced in confined nanoscale magnets with IMA under perpendicular magnetic fields. The chiral magnetic structures depend significantly on the size of the nanostructures. A controlled string of periodic skyrmion states emerges within the central magnetic domain wall, which can be tuned by the steady magnetic fields and the size of the nanostructures. Non-trivial topological states with non-integer topological charges are achieved by tuning the magnetic fields or the sizes of the nanostructures. Importantly, the periodic switching between the trivial and the non-trivial topological configurations is realized using an alternating magnetic field. Our study reveals an important mechanism for controlling novel skyrmion states via nanoconfinement effects and the IMA in magnetic nanostructures, and also provides a new approach for the development of magnetic field-modulated spin nanodevices.
\end{abstract}
\maketitle

\section{Introduction}
To date, extensive research has focused on spin topological textures in bulk and thin film magnetic materials. However, there are relatively limited studies on individual spin texture in geometrically confined magnetic nanostructures with in-plane magnetic anisotropy (IMA), particularly in the presence of steady and alternating magnetic fields. Spintronic devices need to be miniaturized to the nanoscale for practical applications. The recent experiments have revealed that boundary confinement of magnetic nanostructures is crucial because it tilts and twists the magnetization and controls the morphology and formation of skyrmions \cite{jin_control_2017, rohart_skyrmion_2013, hagemeister_2016}, when the size of nanostructures is close to that of the magnetic skyrmion. For instance, it has been demonstrated that geometrically confined skyrmions with a wide range of sizes and ellipticities can be generated in a nanostrip from a helical magnetic state with a distorted edge twist \cite{jin_control_2017}. The study of spin-polarized scanning tunneling microscopy shows that controlling boundary conditions can be utilized to tailor specific properties of skyrmions \cite{hagemeister_2016}. The geometric edges give rise to a natural oscillatory motion for skyrmions in confined structures, resulting in dynamic resonances under a driving current \cite{navau_analytical_2016}. Geometrically tailored skyrmions are achieved based on multilayer confined nanostructures without a magnetic field \cite{tan024064}. Furthermore, magnetic confinement and curling effects in nanostructures are vital in magnetic nanostructures due to symmetry breaking, which can yield new magnetic textures and magnetic phenomena, such as skyrmion tube, leech, toron, and bobber states \cite{tambovtsev_2022, monopole_2018}. The surface boundary confinement effects provide a stabilization mechanism that results in twisted stable skyrmions and surface skyrmions \cite{leonov087202,zhang_robust_2020}. Direct imaging of magnetic field-driven transitions of skyrmion cluster states has been developed in FeGe nanodisks using Lorentz transmission electron microscopy \cite{zhao_direct_2016}. Boundary effects can significantly impact the accurate experimental observation of individual skyrmion \cite{Yu2010ayi, zhao_direct_2016}. These results demonstrate that the geometric confinement effect can be used as an important tool and approach in the manipulation and control of magnetic skyrmions in magnetic nanostructures. However, there is still a lack of theoretical and systematic studies on how nanoconfinement effects control the spin topological states in magnetic nanostructures with the IMA.

Moreover, skyrmion states have been extensively demonstrated and studied in chiral single-layer and thin-film magnets with perpendicular magnetic anisotropy (PMA) \cite{rohart_skyrmion_2013, mulkers_cycloidal_2016, mulkers_2017, Bhattacharya2020, PRB224402, dong_tuning_2023, dongjap_2021}. We have noticed that the current theoretical studies on skyrmions in IMA nanostructures still significantly trail behind the experimental findings. The previous theoretical studies have indicated that stabilized skyrmions are not supported in chiral monolayer magnets, owing to their large IMA, as observed in the Janus VSeTe monolayer \cite{intrinsic_2020,im_dynamics_2019}. It is noted that skyrmions can be found in B20 thin film materials with IMA under an applied perpendicular magnetic field \cite{vousden_2016}. However, the simulation findings are incongruent with the analytical results \cite{Wilson2014,intrinsic_2020,im_dynamics_2019}, and comprehensive theoretical explanations, particularly in confined magnetic nanostructures, are still severely lacking. They also suggested that stable skyrmions can be created in such confined geometry systems. Nevertheless, it has been found that two dimensional (2D) ferromagnetic Janus VSeTe monolayer exhibits strain-controllable high Curie temperature, large valley polarization, and pronounced magnetic crystal anisotropy \cite{guan_2020, guan13988}. Recently, it has been shown that VSeTe is the most promising for electrically controlled spin-orbit torque (SOT) storage, due to its large magnetic anisotropy and strong spin-orbit coupling \cite{smaili_2021}. It is noticed that the skyrmions can be more easily stabilized in 2D magnetic systems with strong spin-orbit coupling \cite{banerjee2014}. However, recent experimental studies of the IMA magnetic system have yielded new insights and findings. The experiment has demonstrated that the interfacial IMA in Co/Pt systems can be improved and controlled \cite{zhou_giant_2024}. The in-plane compass anisotropy induces exotic skyrmion crystals in chiral magnets under out-of-plane magnetic fields \cite{chen_exotic_2016}. Skyrmion crystals can be formed with IMA using in-plane magnetic fields \cite{hayami224418}. The elliptical skyrmions are induced by IMA to control the skyrmion Hall effect \cite{chen174409}. Very recently, it is found that Fe$_5$GeTe$_2$ monolayers can host skyrmions smaller than 10 nm in a perpendicular magnetic field above 2.2 T due to moderate IMA \cite{lil220404}. Furthermore, it has been experimentally demonstrated that nanoscale magnetic skyrmions are stable in thin-film multilayers with the IMA \cite{flacke_robust_2021}. In addition, it has been experimentally found that the power of spin Hall nano-oscillators is significantly amplified by the IMA \cite{montoya_2023}. Skyrmions can be efficiently confined and exhibit steady spin-torque nano-oscillators in a nanodisk due to the IMA-induced modified magnetized textures \cite{luo_ferrom_2024}. These prominent discoveries and outcomes highlighted above open up new possibilities and avenues for exploring the nanoscale skyrmionic states and mechanisms in magnetic nanostructures with the IMA.
 
Furthermore, the manipulation of non-trivial spin textures using alternating magnetic fields is currently garnering significant attention. It has been found that a single skyrmion or skyrmionium can be created by a perpendicular oscillating magnetic field \cite{vigo2020}. The skyrmion resonance modes of skyrmions can be excited and the nanoscale skyrmions can be driven by high-frequency magnetic fields \cite{moon_skyrmion_2016}. The diffusion coefficient of the skyrmion increases exponentially with the increasing alternating magnetic field, thereby providing a technological framework for low-power-Brownian devices \cite{stochastic_2021}. The switching of skyrmion polarity through a transient skyrmionium is achieved by utilizing alternating magnetic fields of very low strength in magnetic hemispherical shells \cite{yan134427}. Additionally, it has been revealed that bimerons can propagate in a specific direction when driven by alternating magnetic fields \cite{shen_dynamics_2020}. Recent studies show that oscillating magnetic fields induce nonlinear dynamics of skyrmions, which results from the coupling of secondary gyrotropic mode with a non-uniform, breathing-like mode \cite{park_emergence_2023}. We believe that these studies open new opportunities for controlling magnetic skyrmions using alternating magnetic fields to develop high-density spin nanodevices.

In this study, we have investigated the non-trivial topological spin textures of IMA-confined nanostructures, based on the micromagnetic analytical model and simulations. Our findings reveal that chiral magnetic tilting occurs at the boundaries, due to the geometrical nanoconfinement effect and the IMA. Magnetic field and size-dependent unique magnetic chiral topological states, such as tilted skyrmions, are uncovered, considering the chiral Dzyaloshinskii-Moriya interaction (DMI). A controlled chain of periodic skyrmion states is generated within the central magnetic domain wall, which the magnetic fields and the size of nanostructures can tune. The non-trivial topological states with non-integer topological charges are observed. Furthermore, the periodic switching between the trivial and non-trivial topological configurations is realized when subjected to an alternating magnetic field. Our findings differ significantly from those in PMA nanostructures and infinite 2D magnetic materials with the IMA. Our study offers a novel approach for the precise controlling and modification of individual skyrmions within confined nanoscale magnetic structures.

\section{Theoretical model and Methodology}

\begin{figure}[htbp]
\centering{}\includegraphics[width=1.0\columnwidth]{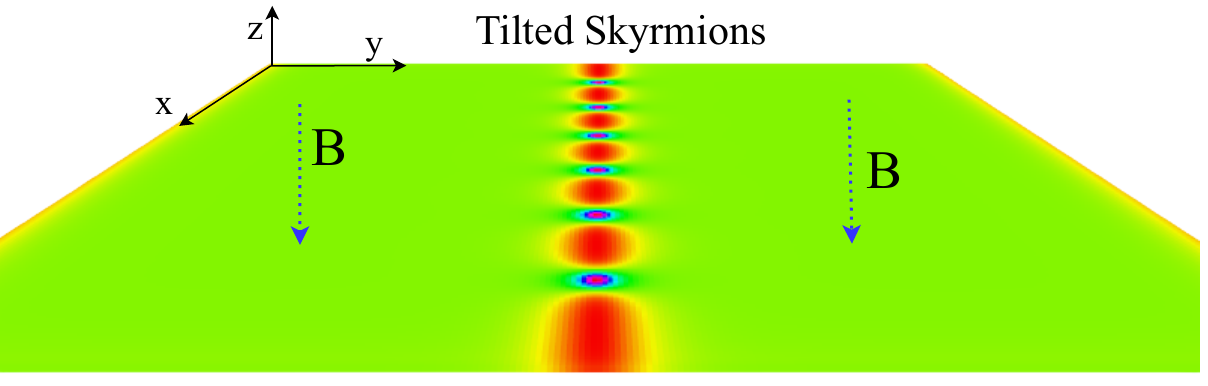}
\caption{Schematic illustration of a confined magnetic nanostructure with IMA subjected to a perpendicular magnetic field B along the negative z-axis direction. The magnetic field and size-dependent tilted skyrmions are produced in the center, whose magnetization at the centers and the edges are tilted, showing the in-plane components of the magnetizations.}
\label{fig1}
\end{figure}

This study focuses on the boundary geometrical confinement effects and the topological magnetic structures of the magnetization vector ${\boldsymbol M}$ in confined nanostructures with the IMA under a perpendicular magnetic field $B$ as shown in Fig. \ref{fig1}. Initially, we utilize a one-dimensional (1D) theoretical model to demonstrate how the magnetization rotates and the explicit effect of magnetic boundary conditions \cite{rohart_skyrmion_2013, mulkers_cycloidal_2016}. The rotation of the magnetization unit vector ${\boldsymbol m}={\boldsymbol M}/M_s$ along the $x$-direction can be uniformly described by the angle $\theta(x)$ referring to the $z$-axis in the $x$-$z$ plane, with the saturation magnetization $M_s$. The perpendicular magnetic field B is in the $z$-axis. Based on the theory of micromagnetism \cite{evans_atomistic_2014}, the total 1D micromagnetic energy can be expressed as
\begin{equation}
\epsilon [\theta(x)] =\int_{x_1}^{x_2}L[\theta(x)] dx, \label{eq1}
\end{equation}
with Lagrangian function
\begin{equation}
L[\theta(x)]=A\bigg(\frac{\partial\theta}{\partial x}\bigg)^2-D\frac{\partial\theta}{\partial x}- K\sin^2\theta-M_s B_z\cos\theta. \label{eq2}
\end{equation} 
$x_1$ and $x_2$ are the corresponding coordinate values at the boundaries of the 1D magnetic nanostructures with the exchange stiffness $A$, the IMA constant $K$ and the interface DMI constant $D$. It is noteworthy that the theoretical model is distinct from the studies in magnetic nanostructures with PMA \cite{rohart_skyrmion_2013, mulkers_cycloidal_2016, mulkers_2017, Bhattacharya2020, PRB224402}, which has been extensively studied. Using the standard Euler's calculus of variations with Eq. \eqref{eq2}, we can obtain
\begin{equation}
\frac{ \partial L }{ \partial \theta }=-2K\cos\theta\sin\theta+M_s B_z\sin\theta,  \label{eq3}
\end{equation} 
and 
\begin{equation}
\frac{ \partial L }{ \partial \theta' }=2A\theta'-D, \label{eq4}
\end{equation} 
with $\theta'=d\theta/dx$. We minimize the total micromagnetic energy to obtain the ground state magnetization distributions through the Euler-Lagrange equation $\frac{ \partial L }{ \partial \theta }-\frac{d}{dx}\Big(\frac{ \partial L }{ \partial \theta' }\Big)=0$ \cite{tai_2024, ma11112238}, thereby we have 
\begin{equation}
\frac{d^2\theta}{dx^2}=\frac{\sin\theta}{\Delta_B^2}-\frac{\sin\theta\cos\theta}{\Delta^2}\quad\mathrm{for~} x_1 < x < x_2, \label{eq5}
\end{equation}
and 
\begin{equation}
\frac{d\theta}{dx}=\frac{1}{\Delta_D}\quad\mathrm{~~~~~~~~~for~}x=x_1\mathrm{~and~}x=x_2. \label{eq6}
\end{equation}
Here $\Delta=\sqrt{A/K}$, $\Delta_B=\sqrt{2A/|M_s B_z|}$ and $\Delta_D=2A/D$, which are all physical quantities in units of length. This suggests that for a 1D magnetic nanoribbon with a finite length $L=|x_2-x_1|$ under a magnetic field, the rotation $\theta(x)$ of the magnetization ${\bf m}$ can be determined using Eq. \eqref{eq5}. As for the Neumann boundary condition, they are provided by Eq. \eqref{eq6} must be fulfilled. We display the results of the magnetization via $m_z=\cos\theta$ for a given $L$ and the difference $\eta=\Delta/\Delta_B=\tau\sqrt{|M_s B_z|/(2K)}$ in Fig. \ref{fig1}. Here $\eta$ denotes the ratio of the perpendicular magnetic field $B$ to the IMA $K$, which varies with the magnetic fields. $\tau=\pm$ denotes the direction of the perpendicular magnetic field along the z-axis.

Next, we examine 2D monolayer magnetic nanostructures with the IMA under a perpendicular magnetic field $B$ as shown in Fig. \ref{fig1}. As a result, the magnetization rotation of the magnetic non-trivial structures that we focus on can be described by the angle $\theta(r)$ with the $z$-axis in the polar coordinate system if skyrmionic states are formed in the center \cite{butenko_2010}. $r$ is the radius in polar coordinates. Thus, the total 2D micromagnetic energy of such nanostructures can be written as follows
\begin{equation}
E[\theta(r)]=2\pi d_0 \int_0^R L[\theta(r)] dr, \label{eq10}
\end{equation}
with $d_0$ being the effective thickness and the radius $R$ of a single-layer magnetic nanodisk, and the 2D Lagrangian function 
\begin{equation}\begin{aligned}
L[\theta(r)]&=Ar\bigg[\bigg(\frac{d\theta}{dr}\bigg)^2+\frac{\sin^2\theta}{r^2}\bigg]-Dr\biggl[\frac{d\theta}{dr}+\frac{\cos\theta\sin\theta}{r}\biggr]
\\&-Kr\sin^2\theta-M_s B_z r\cos\theta.   \label{eq11}
\end{aligned}\end{equation}
As a result, we have the two equations with the 2D Lagrangian function, which are
\begin{equation}
\frac{ \partial L }{ \partial \theta }=\Big(\frac{A}{r}-Kr\Big)\sin2\theta-D\cos2\theta+M_s B_z r\sin\theta, \label{eq12}
\end{equation} 
and 
\begin{equation}
\frac{ \partial L }{ \partial \theta' }=2Ar\theta'-Dr, \label{eq13}
\end{equation} 
with $\theta'=d\theta/dr$. Based on the Euler–Lagrange equation for minimizing the total 2D micromagnetic energy \cite{tai_2024, ma11112238}, it leads to 
\begin{equation}
\theta''+ \frac{\theta'}{r}-\frac{\sin^2\theta}{r\Delta_D}-\Big(\frac{1}{2r^2}-\frac{1}{2\Delta^2}\Big)\sin(2\theta)-\frac{\eta^2\sin\theta}{\tau\Delta^2}=0,  \label{eq14}
\end{equation} 
and 
\begin{equation}
\frac{d\theta}{dr}=\frac{1}{\Delta_D}\quad\mathrm{~~~~~~~~~~~~~~~~for~}r=R.  \label{eq15}
\end{equation}
We have numerically solved the partial differential Eq. \eqref{eq14}, using the Neumann boundary condition Eq. \eqref{eq15} \cite{kobayashi_2013, mulkers_cycloidal_2016}. This is different from the Dirichlet boundary conditions commonly used for skyrmions \cite{Wilson2014, ma11112238, jiang013229}. To achieve a stable and reliable solution, we segment the entire interval into several sub-intervals and conduct multiple attempts to solve and fit the equation numerically, ensuring that the specified boundary conditions are met throughout the process \cite{lvarez_2008}. The magnetization profiles $m_z$ of the formed skyrmionic states are presented with the radius $R=32$ nm and $R=64$ nm of the nanodisks at the different magnetic fields B in Fig. \ref{fig2}, respectively.

Furthermore, we engage in the 2D magnetic nanostructures with IMA and perform high-throughput calculations through micromagnetic simulations, to find stable magnetic configurations and ensure the accuracy of magnetic theoretical analysis \cite{Vansteenkiste2014, joos_tutorial_2023}. The theoretical model and results are equally applicable to other thin films or monolayer chiral nanomagnets with the IMA in the presence of a perpendicular magnetic field. The total micromagnetic energy of the systems, $E_{tot} = E_{ex}+ E _{DMI} + E_{IMA} + E_{Z}$, includes magnetic exchange energy, antisymmetric exchange DMI energy, IMA energy, Zeeman energy induced by an applied field B, respectively. The magnetization unit vector ${\bf m}$, which characterizes the instantaneous magnetic state, is determined by solving the Landau-Lifshitz-Gilbert (LLG) equation, namely $\partial_t \boldsymbol{m}=-\gamma\boldsymbol{m}\times\boldsymbol{H}_{\mathrm{eff}}+\alpha\boldsymbol{m}\times\partial_t\boldsymbol{m}$ \cite{fuming2023, lakshmanan_2011}. The gyromagnetic ratio $\gamma =2.21\times 10^5$ m$\cdot$A$\cdot$s$^{-1}$ and $\alpha$ refers to the phenomenological Gilbert damping parameter \cite{gilbert_2004}. The effective field $H_{{\mathrm{eff}}}=-(M_s\mu_{0})^{-1}\delta E_{tot}/\delta\boldsymbol{m}$ is the functional derivative of the total energy function with vacuum magnetic permeability $\mu_0$. The continuum micromagnetic model is applied to compute and obtain the magnetization field $\boldsymbol{M} = M_{s}\cdot\boldsymbol{m}$ of arbitrary structures \cite{mulkers_2017}. 

In this work, we consider a 2D magnetic nanostructure (see Fig. \ref{fig1}), for example, Janus VeSeTe with the saturation magnetization $M_s=5.357\times 10^5$ A/m, the exchange stiffness $A=3.9\times 10^{-12}$ J/m, the IMA constant $K_u=4.42\times 10^6$ J/m$^3$, and the interfacial DMI constant $D=2.0\times10^{-3}$ J/m$^2$, which can be calculated by the first-principles calculations \cite{intrinsic_2020}. Janus magnetic VeSeTe has exhibited a high Curie temperature and possessed strong IMA \cite{guan_2020, guan13988}. The fixed Gilbert damping parameters $\alpha=0.02$ and $d_0=0.4$ nm are used for VSeTe. The effective IMA $K=K_u-\mu_0M_s^2/2$ includes the demagnetizing field energy in the local approximation. We have conducted a detailed analysis of magnetic nanostructures with varying sizes and shapes in the presence of steady and alternating magnetic fields perpendicular to the structures. This procedure has been aided by the massive GPU-based computing program Mumax3 \cite{Vansteenkiste2014}. In addition, we have thoroughly examined the magnetic structure of the perpendicular magnetic fields at different frequencies.

\section{Results and Discussions}
\begin{figure}[htbp]
\centering{}\includegraphics[width=0.8\columnwidth]{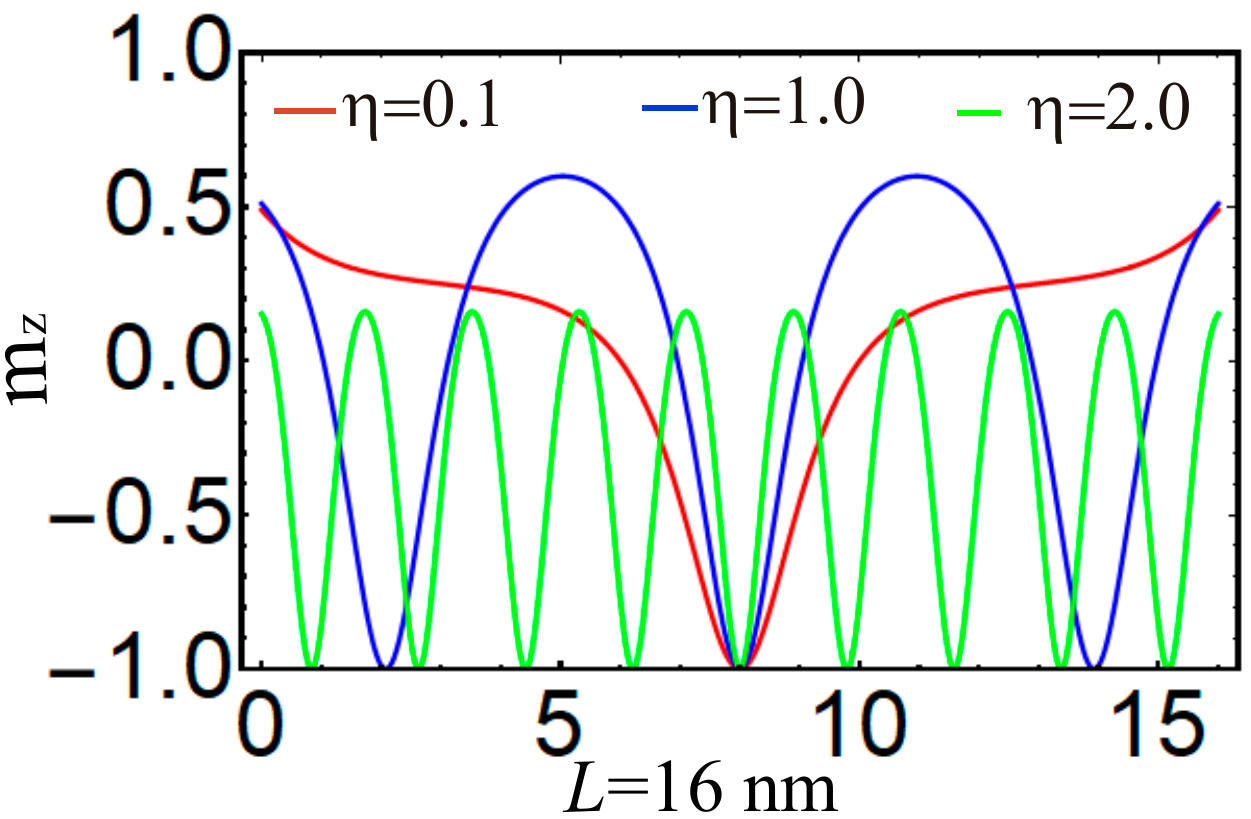}
\caption{Variation of the magnetization across $m_z$ in a 1D nanostructure with the length $L$ for the different steady perpendicular magnetic fields B, given by Eqs. \eqref{eq5} and \eqref{eq6}. }
\label{fig2}
\end{figure}
Our study focuses on the 2D confined nanostructures with IMA in the presence of a perpendicular magnetic field, which is significantly different from the previous results on the PMA \cite{rohart_skyrmion_2013, mulkers_cycloidal_2016}. In Fig. \ref{fig2}, we show the magnetization profile $m_z$ in 1D VTeSe nanostructures with IMA and the length $L=16$ nm in the different strengths of perpendicular steady magnetic fields B, obtained by Eqs. \eqref{eq5} and \eqref{eq6}. This indicates that symmetric spin cycloid states are formed, the period of which depends strongly on the nanosize $L$ and the perpendicular magnetic field. The result is different from that observed in ultrathin film nanostructures with PMA \cite{rohart_skyrmion_2013, mulkers_cycloidal_2016}. The size of the period decreases as the strength of the steady external magnetic field increases. The magnetic boundary conditions are the same in both cases, namely PMA and IMA in our study [see Eq. \eqref{eq6}], but the rotations of the magnetic magnetization are different [see Eq. \eqref{eq5}]. Fig. \ref{fig2} shows that the nanoconfinement effect can significantly control the magnetic structure. At the edges, the magnetization is significantly tilted since the magnetization z-component $m_z$ is less than 1 ($m_z < 1$), i.e., $\theta \neq 0$. The IMA aligns the spins along the x-axis, while the perpendicular magnetic field tends to align in the z-axis. Note that the PMA causes the spin to favor the z-axis direction. Cycloidal states are formed for PMA only when $D > D_c = 4\sqrt{AK}/\pi$, but the condition is not required for IMA. Taking the chiral DMI and IMA into account, this leads to the formation of the collinear cycloid states. Recently, a similar result has been discovered experimentally that spin cycloids can also originate from the magnetoelastic anisotropy resulting from the substrate epitaxy \cite{meisenheimer_2024}. 

\begin{figure}[htbp]
\centering{}\includegraphics[width=0.8\columnwidth]{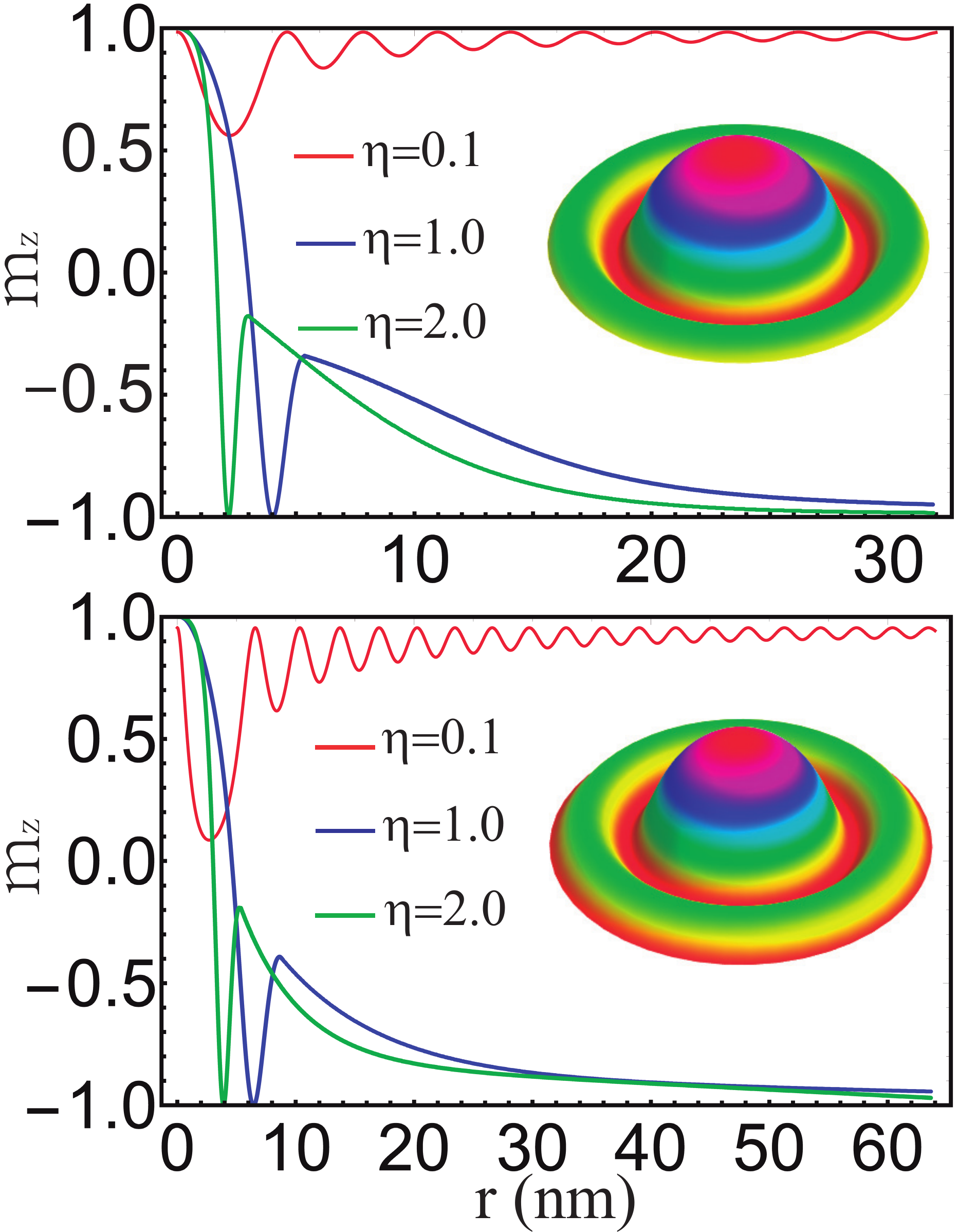}
\caption{Magnetization profile across $m_z$ of skyrmion states as a function of the radius $r$ for the radius $R=32$ nm (upper) and $R=64$ nm (lower) of the 2D nanodisk at the different steady magnetic fields B, obtained by numerically solving Eqs. \eqref{eq14} and \eqref{eq15}. The inserts are the corresponding three-dimensional contour maps of $m_z$ for $\eta=1.0$ respectively.}
\label{fig3}
\end{figure}

\begin{figure}[htbp]
\centering{}\includegraphics[width=1.0\columnwidth]{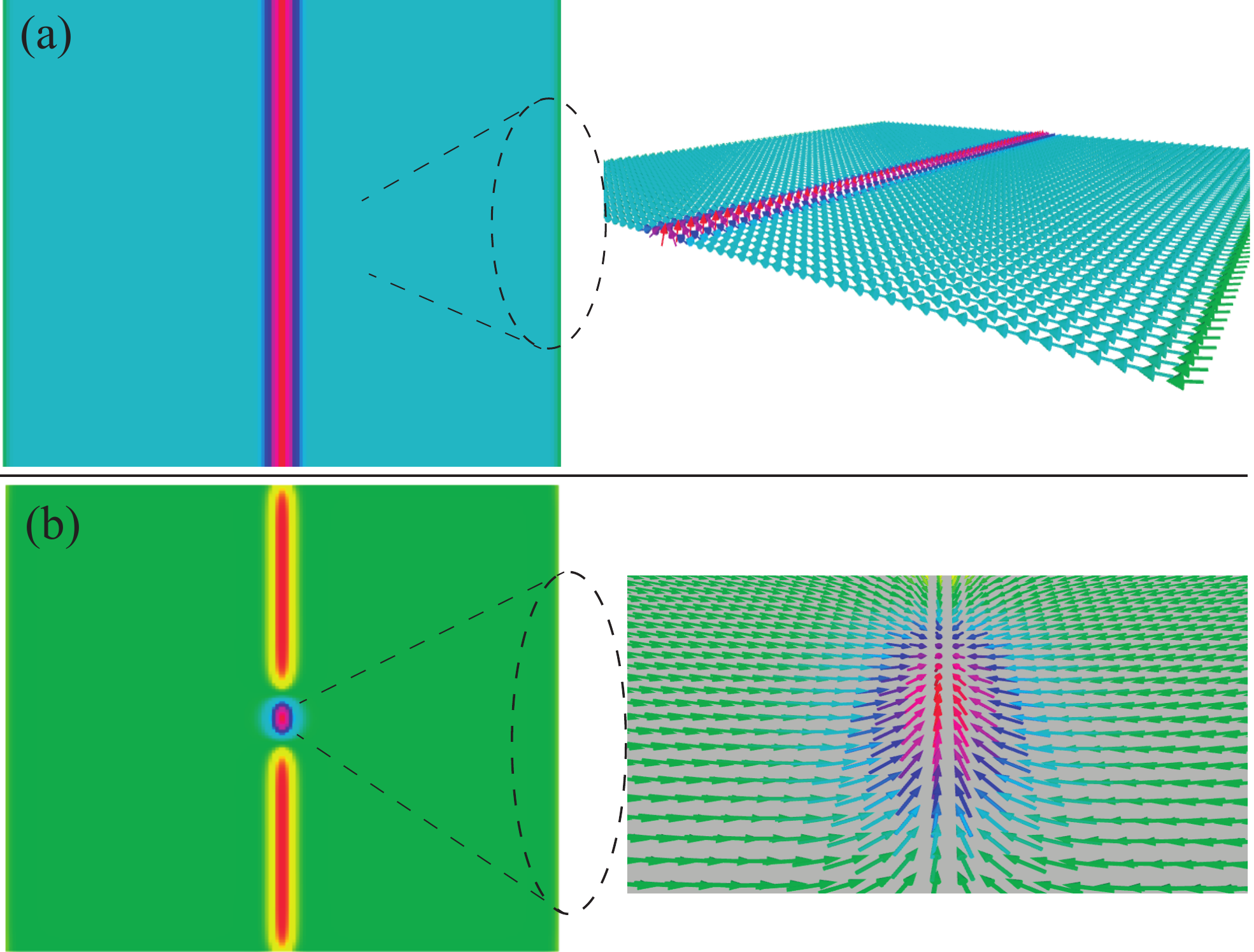}
\caption{The magnetization profiles of the simulations for the $64\times64$ nm magnetic nano-square structure under the perpendicular magnetic field (a) $B=0$ and (b) $B=-4.7$ T.}
\label{fig4}
\end{figure}

We show the magnetization profile cross $m_z$ and the counter maps (insert) of the central skyrmion states for the radiuses $R=32$ nm (top) and $R=64$ nm (below) of the 2D nanodisks at the different steady magnetic fields B by solving the Eqs. \eqref{eq14} and \eqref{eq15} in Fig. \ref{fig3}. In a relatively weak magnetic field, the spins cannot flip $180^{\circ}$, but a $180^{\circ}$ spin reversal of the nanomagnetic structures can be realized and the novel non-trivial skyrmionic states with very small sizes are tailored in the center under a strong magnetic field. Our theoretical results are in line with the earlier experimental findings \cite{hayami224418,chen174409,lil220404,flacke_robust_2021}. The skyrmionic states are dominated by the external magnetic field $B$ and the size of the nanodisks, for example, the radius $R$. The Neumann boundary conditions of 2D are the same as those for 1D. These results for 2D nanodisks with IMA are very different from those in 2D nanodisks with PMA \cite{rohart_skyrmion_2013, mulkers_cycloidal_2016}. We find that the spin exhibits rapid rotation behavior and that these rotations become more pronounced as the magnetic field becomes more intense in magnetic nanodiscs with IMA. The former leads to an in-plane alignment of spins, while the latter gives rise to a perpendicular alignment of spins, and the DMI produces helical rotations. Due to symmetry breaking, the confinement effect and the chiral DMI lead to the chiral magnetic tilting at both edges and centers. As a result, the spin textures get titled and twisted (see also Eq. \ref{eq15}). The rapid rotations of the spin are induced in the presence of magnetic fields in Fig. \ref{fig3}, as the twisting is progressively transmitted towards the center. The confinement effect can also lead to the creation of two domains with opposite curls to reduce the dipolar energy of the nanosystems, as shown in the simulation results in Fig. \ref{fig4}. 

Without an external magnetic field B, two magnetic domains with opposite chirality form in the nanostructures [see Fig. \ref{fig4} (a)]. The transition between the two magnetic domains occurs through a domain wall in the center. The magnetic structure flips within the domain wall at the center of such nanostructures, causing the magnetization structure to curl at both edges. In the magnetic nano-square structures, the magnetic configurations and the tilting at the edges are demonstrated in Fig. \ref{fig4}. Although the presence of domain walls increases the total magnetic energy, the twisting of the magnetic structure reduces the static magnetic energy and the stray field of the entire magnetic system due to the boundary confinement. However, the magnetizations inside the centers of domain walls persist in rotating and twisting, giving rise to non-trivial topological states like magnetic skyrmions in the presence of a perpendicular magnetic field, for example, $B=-4.7$ T in Fig. \ref{fig4} (b). The physical reason for this is that the magnetic field can change the tilted state at the boundary, and the magnetization of the domain wall at the center must also change. To realize the lowest total energy state, the domain walls in the center twist to produce a chiral skyrmion state, taking into account the DMI. We find that a perpendicular magnetic field can effectively control non-trivial topological textures in domain walls. Meanwhile, the simulation results shown in Fig. \ref{fig4} are in line with the findings of the theoretical analysis in Fig. \ref{fig3}.  

\begin{figure}[htbp]
\centering{}\includegraphics[width=1.0\columnwidth]{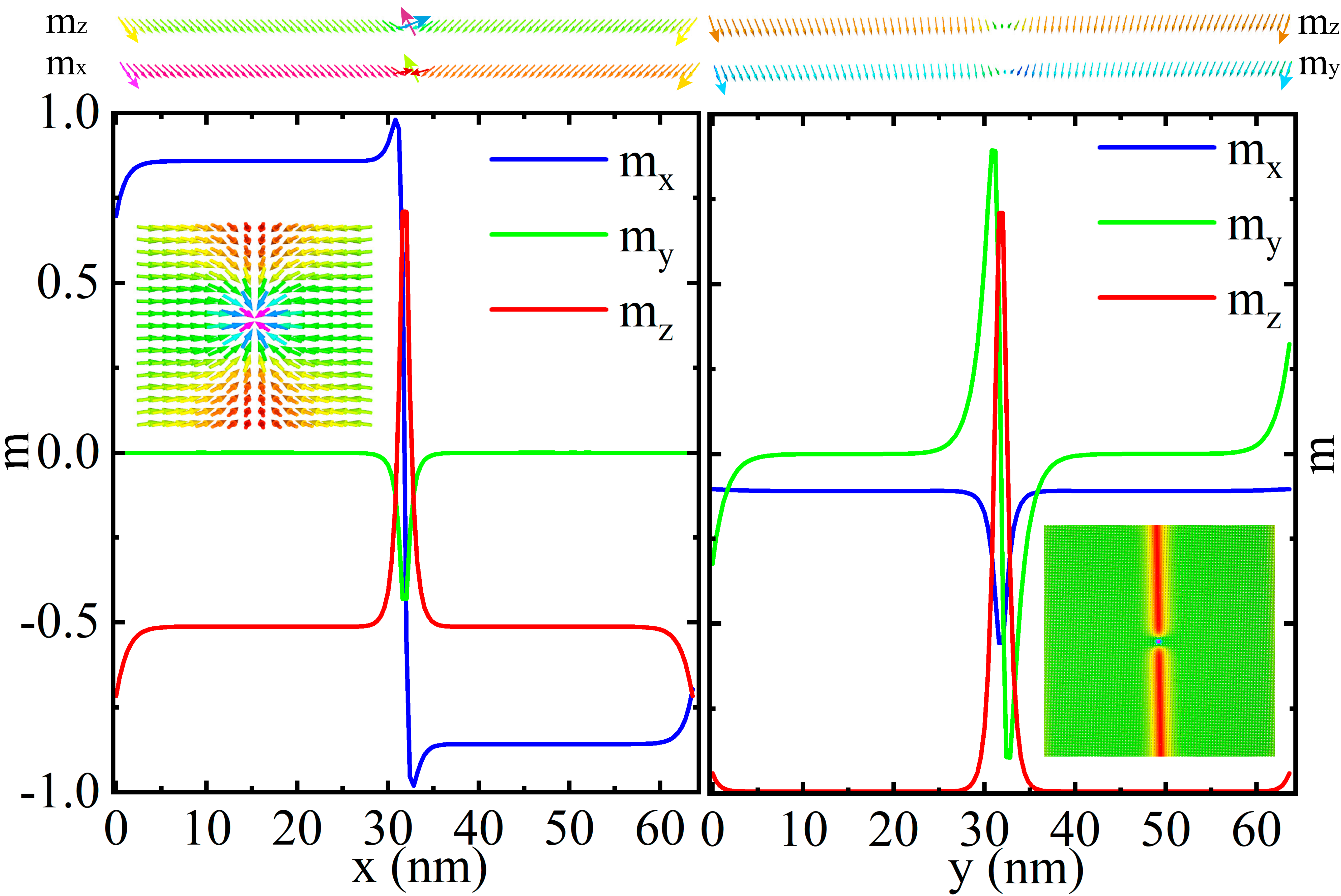}
\caption{The magnetization rotations (above) and the profiles (below) of the non-trivial skyrmion at the center for the $64\times64$ nm magnetic nano-square at $B=-4.7$ T along the $x$ (left) and $y$ (right) directions respectively. The inserts are a detailed view of the moment distribution of the skyrmion (left) at the center and the whole magnetic configuration (right). The magnetizations at the edges and the center are highlighted and the magnetizations rotate inwards at the center (right). In both the subfigures, the spins constantly rotate in planes that are perpendicular to each other.}
\label{fig5}
\end{figure}

In Fig. \ref{fig5}, we show the magnetization rotations and the profiles of the non-trivial skyrmion at the center for the $64\times64$ nm magnetic nano-square at $B=-4.7$ T in the $x$ and $y$ directions, respectively, to comprehend the skyrmion structures and the magnetization rotations. There are noticeable tiltings of the magnetizations, especially at the center and the edge (see the tops in Fig. \ref{fig5}). Moreover, the tiltings at the centers and the edges lead to $m_z < 1$ and the magnetization tiltings at the boundary remain the chirality (see the subfigures below in Fig. \ref{fig5}). The spin has completely reversed from the center to the edge, despite the presence of the tilting. Please refer to the tops of Fig. \ref{fig5}, where the rotations of the magnetization, e.g. $m_x, m_y$, and $m_z$ are each shown in detail, respectively. It is found that the successive rotations of the tilted spin are observed within the center of the magnetic domain walls, resulting in a non-trivial topological structure, such as novel tilted skyrmions with the topological charge $Q=1$ in Fig. \ref{fig5}. For example, the tilting and rotation of $m_x$ and $m_z$ are shown in the left subfigure of Fig. \ref{fig5}, while the tilting and rotation of $m_y$ and $m_z$ are exhibited in the right subfigure of Fig. \ref{fig5}, due to the tilting of spins in the boundary. The competition between the perpendicular magnetic field and the IMA is essential for tilting, particularly within the symmetric boundary confinement effects. The results differ significantly from those for magnetic skyrmions in PMA nanostructures and infinite 2D magnetic materials. Notably, the magnetizations in the center and at the boundary of the latter two are not tilted, but the other skyrmionic states are yielding, which we have studied \cite{dong_tuning_2023,dongjap_2021}.

\begin{figure}[htbp]
\centering{}\includegraphics[width=0.8\columnwidth]{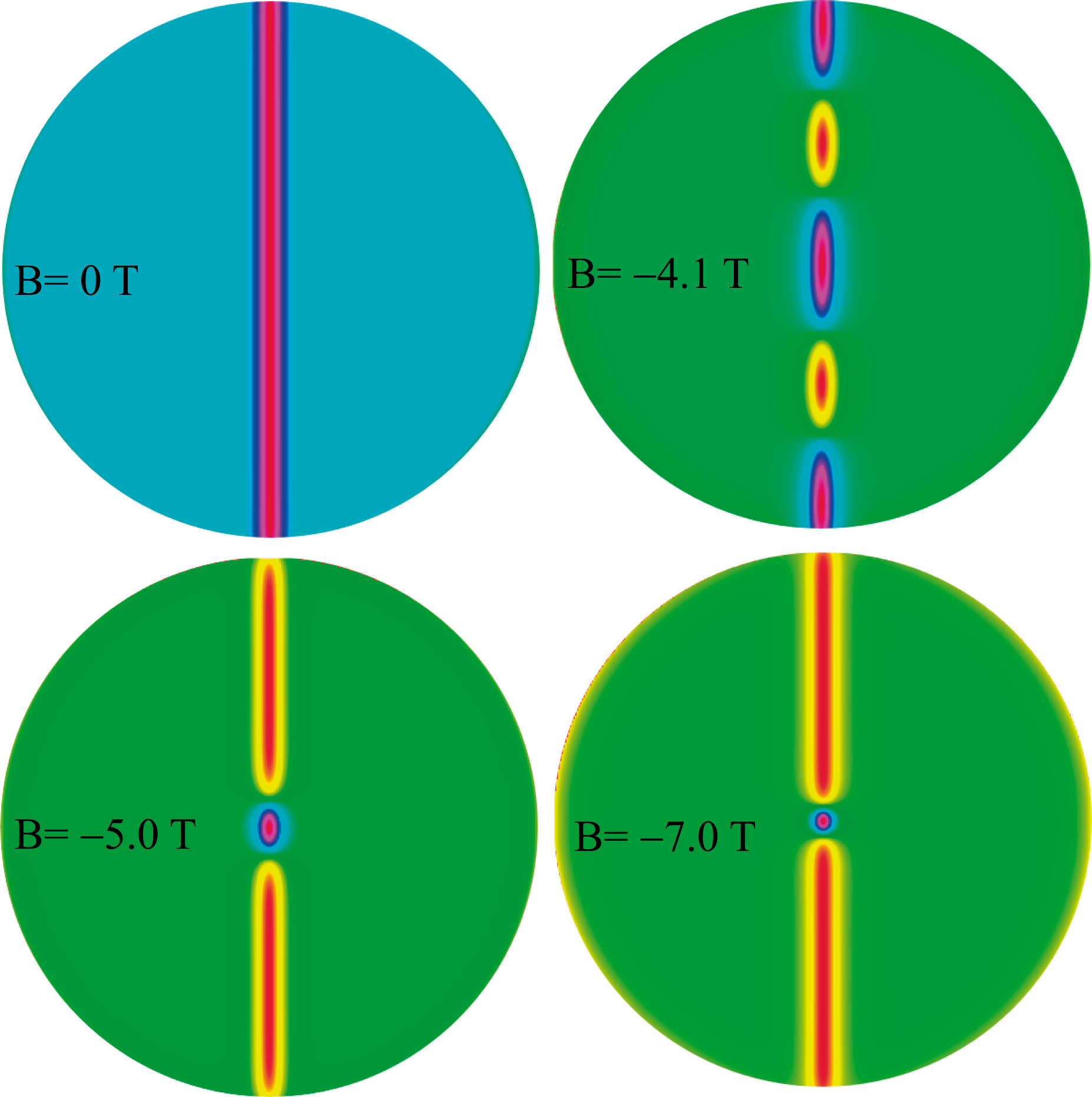}
\caption{The topological configurations of the nanodisk with the diameter $64$ nm at the different perpendicular magnetic fields B. As the reverse magnetic field B increases, the topological charges of the four spin textures are 0, 2, 1.2, and 1, respectively.}
\label{fig6}
\end{figure}

Tilted skyrmions are extremely sensitive to external magnetic fields, as shown in Fig. \ref{fig6}. 
The topological spin textures of the nanodisk with a diameter 64 nm at the different perpendicular magnetic fields B are presented. In confined nanostructures with IMA, the magnetic tilts at the boundary always exist irrespective of the presence or absence of a magnetic field. These results are similar to those in nanosquares, such as Figs. \ref{fig4} and \ref{fig5}, which also agree with the results of our theoretical analysis in Fig. \ref{fig3}. IMA tends to form an in-plane distribution of magnetizations, whereas a perpendicular magnetic field can produce a perpendicular distribution of magnetizations, and DMI results in chiral spiral textures. Non-trivial spin states are ultimately condensed as a result of the competition of the three interactions together. The geometrically nanoconfined effect around the nanodisks induces spin tilting that propagates progressively, and tilted skyrmion states are formed at the center. The magnetic field can change the rotations and the tilting of the magnetizations, including the magnetizations at the center and the boundary. The tilted skyrmions yield only in the central part of the domain walls and can be tuned by the external magnetic fields because the magnetizations at the edges all remain the same tilting and the tilted magnetizations at the edges can progressively flip towards the center. Our study indicates that the topological spin textures, such as the skyrmions, in magnetic nanoconfined structures exhibit unique properties compared to the skyrmions in general magnetic systems, which opens up new opportunities for further control of geometrically-confined nano-skyrmions.

\begin{figure}[htbp]
\centering{}\includegraphics[width=1.0\columnwidth]{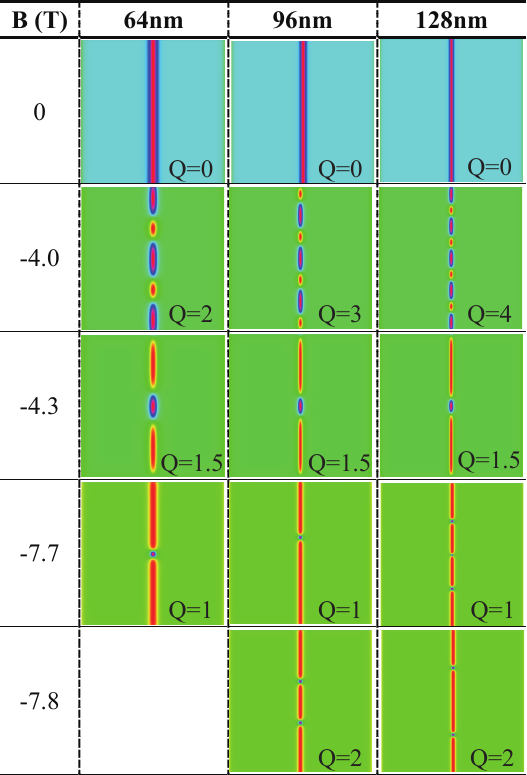}
\caption{The magnetization configurations and the topological charges $Q$ in VSeTe Janus monolayer of the nano-square for $64\times64$ nm (left), $96\times96$ nm (center) and $128\times128$ nm (right) within the different magnetic fields B, respectively.}
\label{fig7}

\end{figure}
\begin{figure}[htbp]
\centering{}\includegraphics[width=1.0\columnwidth]{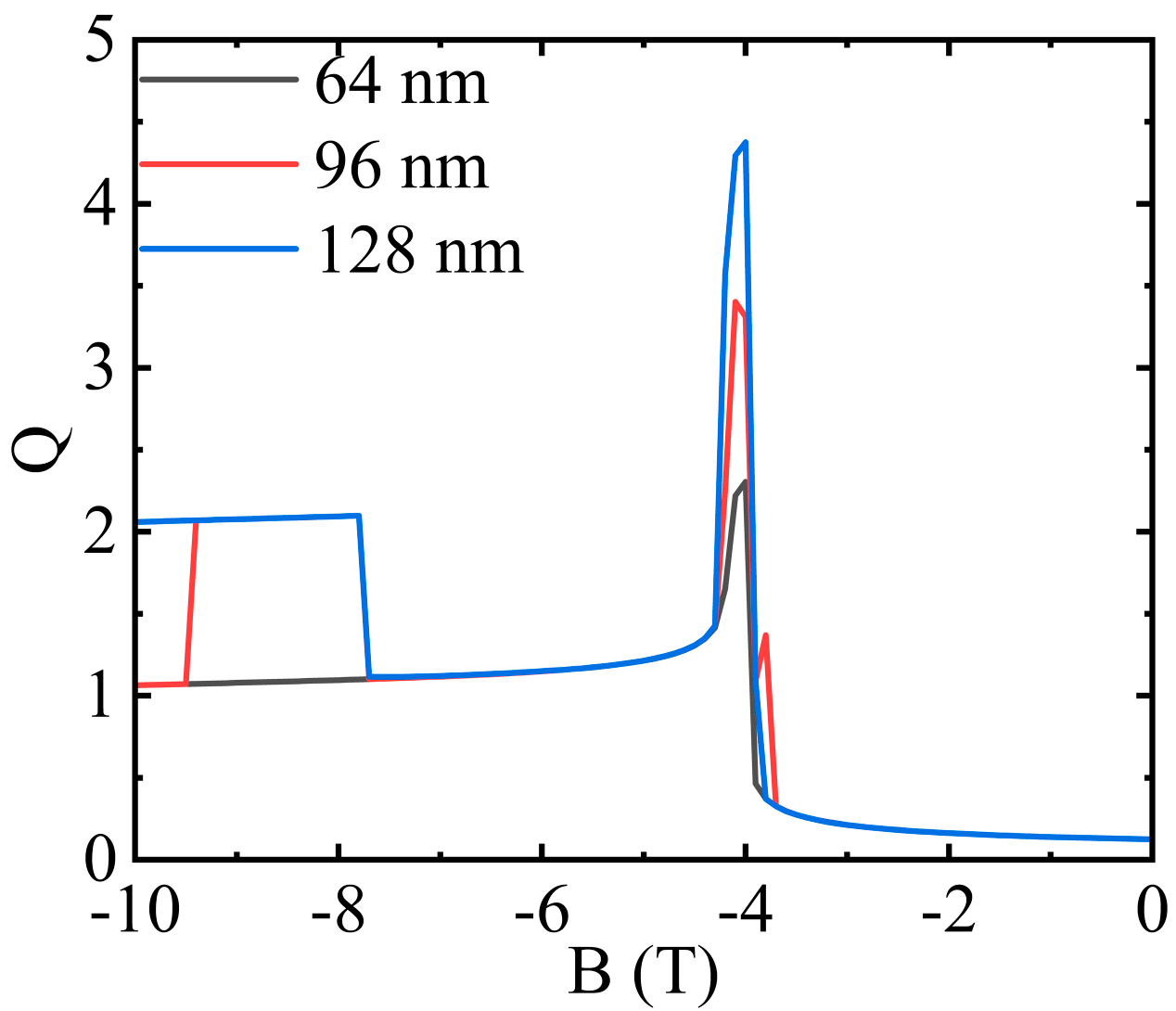}
\caption{The topological charges as the function of the perpendicular steady magnetic field B for the different nano sizes as mentioned in Fig. \ref{fig7}. }
\label{fig8}
\end{figure}

We present the magnetization configurations and the topological charges $Q$ for the different nanosizes and magnetic fields in Fig. \ref{fig7}. It shows the magnetic field and the size-dependent unique magnetic chiral topological states in magnetic nanostructures with IMA. Without a magnetic field $(B=0)$, two trivial ferromagnetic states ($Q=0$) with opposite tilts, for example, left tilting and right tilting in Fig. \ref{fig5}, are connected by a domain wall. It should be noticed that the chiral state in the center possesses a topological charge of $Q = -1$, while the topological charges of the chiral states above and below are both $Q = +1$ for B = $-7.7$ T with the 128 $\times$ 128 nm size. Consequently, the total topological charge for the whole nano-system is $Q = +1$. As the reverse magnetic field increases further, the small chiral structure in the middle is annihilated, leaving only the two skyrmion states, which yield a total topological charge of $Q = 2$ together. As the reverse magnetic field increases, a series of non-trivial skyrmion states with the different $Q$ are generated by twisting the magnetizations in the central domain wall. Furthermore, the skyrmion state $Q$ varies with nanosize due to distinct geometrically nano-confined effects, resulting in a different spin rotation at the center. The strings of tilted skyrmions can only be generated in the central domain wall, not elsewhere in the nanostructures, while the other parts of the confined nanostructures remain in the ferromagnetic states. As the value of IMA $K$ value decreases, the strength of the reverse magnetic field B necessary to generate tilted skyrmions at the center correspondingly decreases.

Fig. \ref{fig8} shows the corresponding topological charge $Q$ as a function of the perpendicular magnetic field B at the different nano sizes. The simulations demonstrate that the magnetic field can trigger the formation of skyrmions. However, if the magnetic field becomes sufficiently strong, the skyrmions collapse and are destroyed. Transitions of the spin textures are not smooth, due to the discontinuous nature of the topological charges. Under the same magnetic field strength B, magnetic nanostructures with different sizes exhibit the different topological charges $Q$. We detail the topological magnetic structures and the transition process as the magnetic field strength varies from 0 to $-10$ T in a $64\times64$ nm nanostructure in Video 1 of the Supplementary Material \cite{supply}. Furthermore, superconducting magnets are remarkable devices that can produce intensity-tuning strong magnetic fields when cooled to extremely low temperatures \cite{Dong21}. Finally, our theoretical simulation results can be verified by state-of-the-art electron holographic imaging \cite{jin_control_2017}, spin-polarized scanning tunneling microscopy \cite{hagemeister_2016} and Lorentz transmission electron microscopy \cite{zhao_direct_2016}.

\begin{figure}[htbp]
\centering{}\includegraphics[width=1.0\columnwidth]{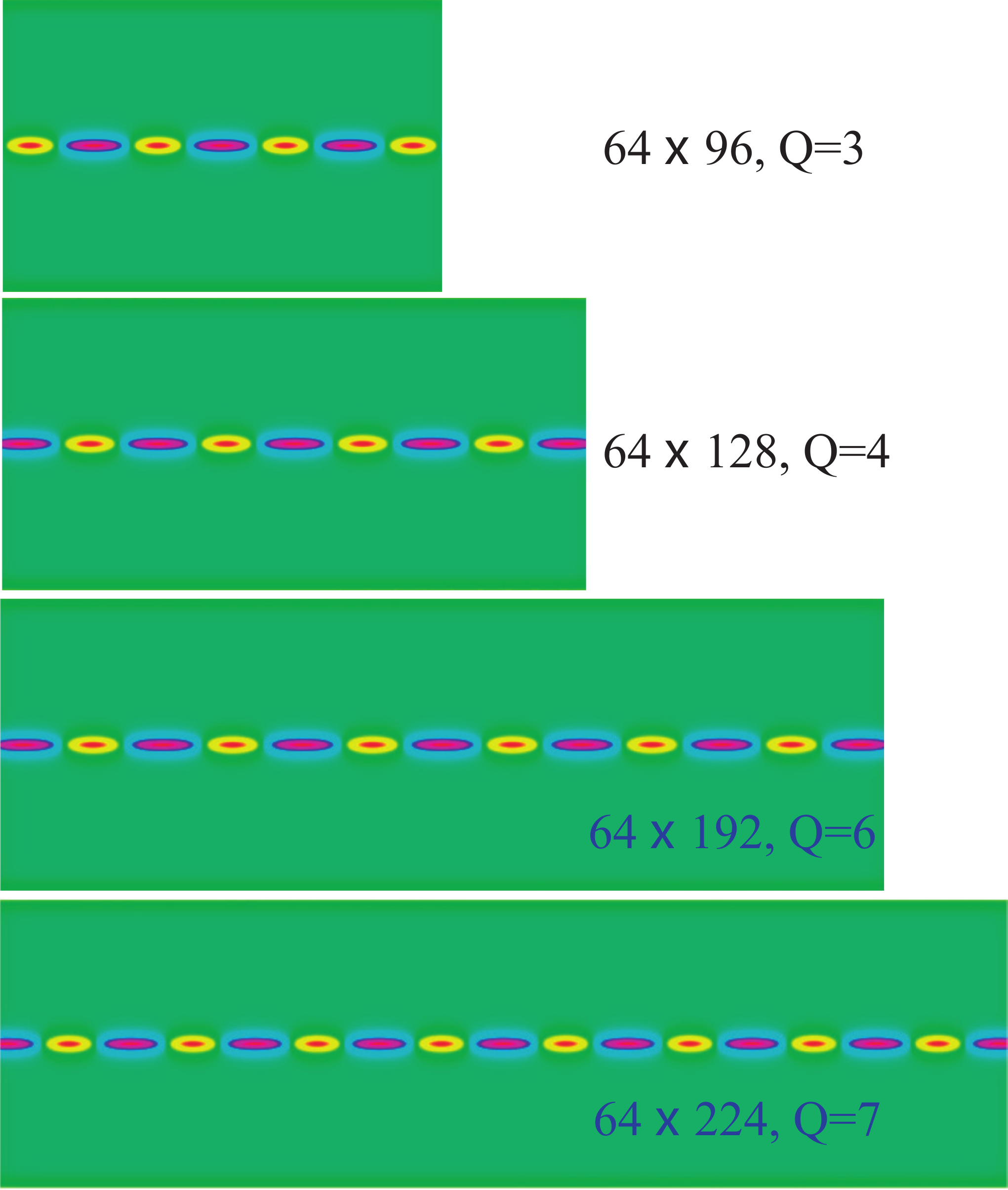}
\caption{The topological skyrmions in the nanoribbon with the fixed width $64$ nm for the different lengths at the fixed perpendicular magnetic field $B=-4.1$ T. }
\label{fig9}
\end{figure}

\begin{figure}[htbp]
\centering{}\includegraphics[width=1.0\columnwidth]{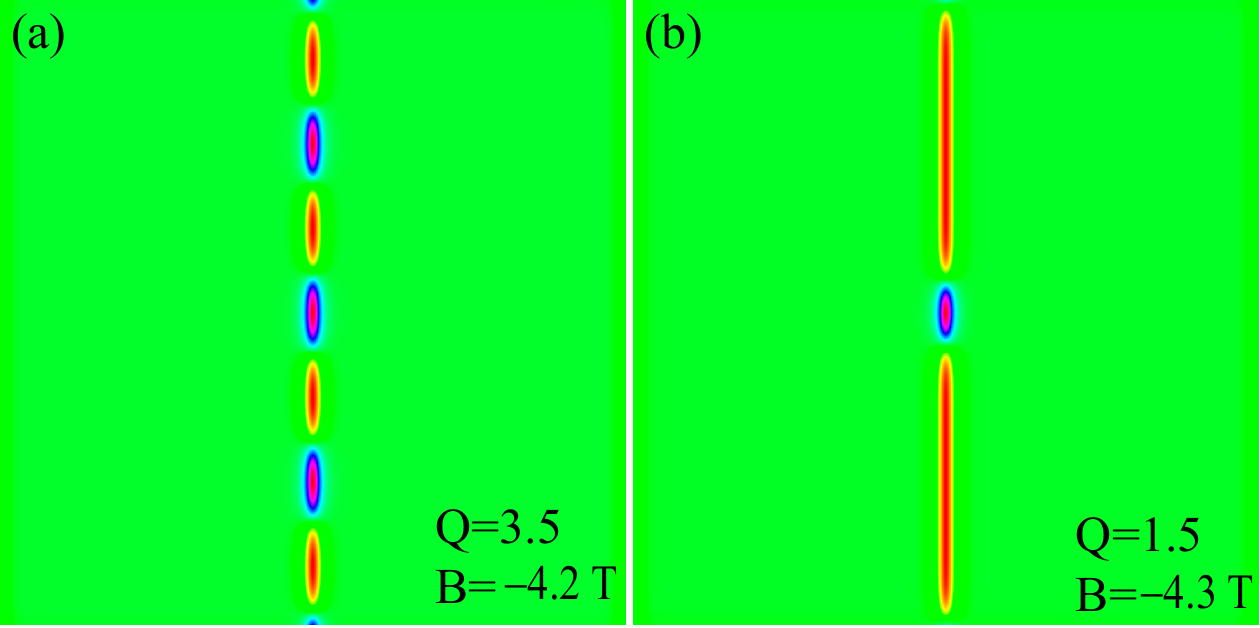}
\caption{The non-trivial topological configurations with non-integer topological charges in the $128\times128$ nm single layer for the different steady magnetic fields B, respectively.}
\label{fig10}
\end{figure}

The development of spin storage devices depends on the controlled generation of precise numbers of magnetic skyrmions. State-of-the-art electron holographic imaging has been developed to directly visualize the skyrmion strings in a wedge-shaped FeGe nanostripe \cite{jin_control_2017}. We have found that the controlled number of magnetic skyrmions in such nanoribbons can be designed and realized by tuning the size (see Fig. \ref{fig9}). In Fig. \ref{fig9}, regularly arranged chains of skyrmions are obtained and all skyrmions are aligned at the center of the magnetic domain wall in the different sizes of the nanostripe. When the nanosizes of magnetic structures are fixed, different skyrmion states with the changing of non-integer topological charges can be obtained by tuning the magnetic field, and vice versa. The skyrmion chain and the distorted edge spins in FeGe nanostripes have been experimentally observed by using high-resolution Lorentz transmission electron microscopy under an applied magnetic field \cite{du_edge_2015}. Our study can also be used to explain and understand these experimental findings. More significantly, we can precisely produce a single or a string of non-trivial topological states with non-integer topological charges, as shown in Fig. \ref{fig10}. The non-integer topological charge arises due to the geometric confinement effects. The magnetization at the edges is twisted and is progressively reversed towards the center induced by magnetic fields. However, a 180$^\circ$ rotation is not achievable in geometrically-confined nanostructures. This ultimately leads to the fact that a skyrmion with an integer topological charge cannot be formed in nanostructures. This formation mechanism is very different from that found for merons and vortices with $Q=0.5$. In contrast to the skyrmion states in PMA nanostructures, these results are completely different. We have shown in earlier work that a variety of non-trivial topological spin textures can form, including flower- and windmill-like skyrmions, skyrmioniums, skyrmion bags, and so on, due to the geometric effects of nanoconfinement in PMA magnetic nanostructures \cite{dong_tuning_2023,dongjap_2021}. This suggests that nanomagnetic IMA structures offer more flexibility in manipulating skyrmions. This also implies that we can design and tailor the skyrmion states in the 2D magnetic nanostructures as needed, which is very favorable for using skyrmions as an information storage nano-unit of devices. 

Furthermore, in 2D or thin-film magnetic materials and liquid crystal ferromagnets, typical skyrmion gas or lattices and the other chiral states form without the edge effects, and external fields can tune these chiral states \cite{behera_magnetic_2019, yu_aggregation_2018}. The geometrically-confined skyrmions with a wide range of sizes and ellipticities can be experimentally realized in magnetic nanostripes \cite{jin_control_2017, zhao_direct_2016}. Very recently, the hopfion rings have been observed in confined nanoplates \cite{hopfion_2023}. In our study, we demonstrate that nano-geometric confinement effects can play a key role, leading to the formation of tilted individual skyrmions or chains of skyrmions. In particular, the skyrmions with non-integer topological charges can be tailored effectively. Recently, the magnetization tiltings in FeGe nanospheres with varying sizes and applied fields have been inadequately explained \cite{pathak_2021}. Our theoretical model elucidates this phenomenon, offering valuable insights and explanations.
\begin{figure}[htbp]
\centering{}\includegraphics[width=0.99\columnwidth]{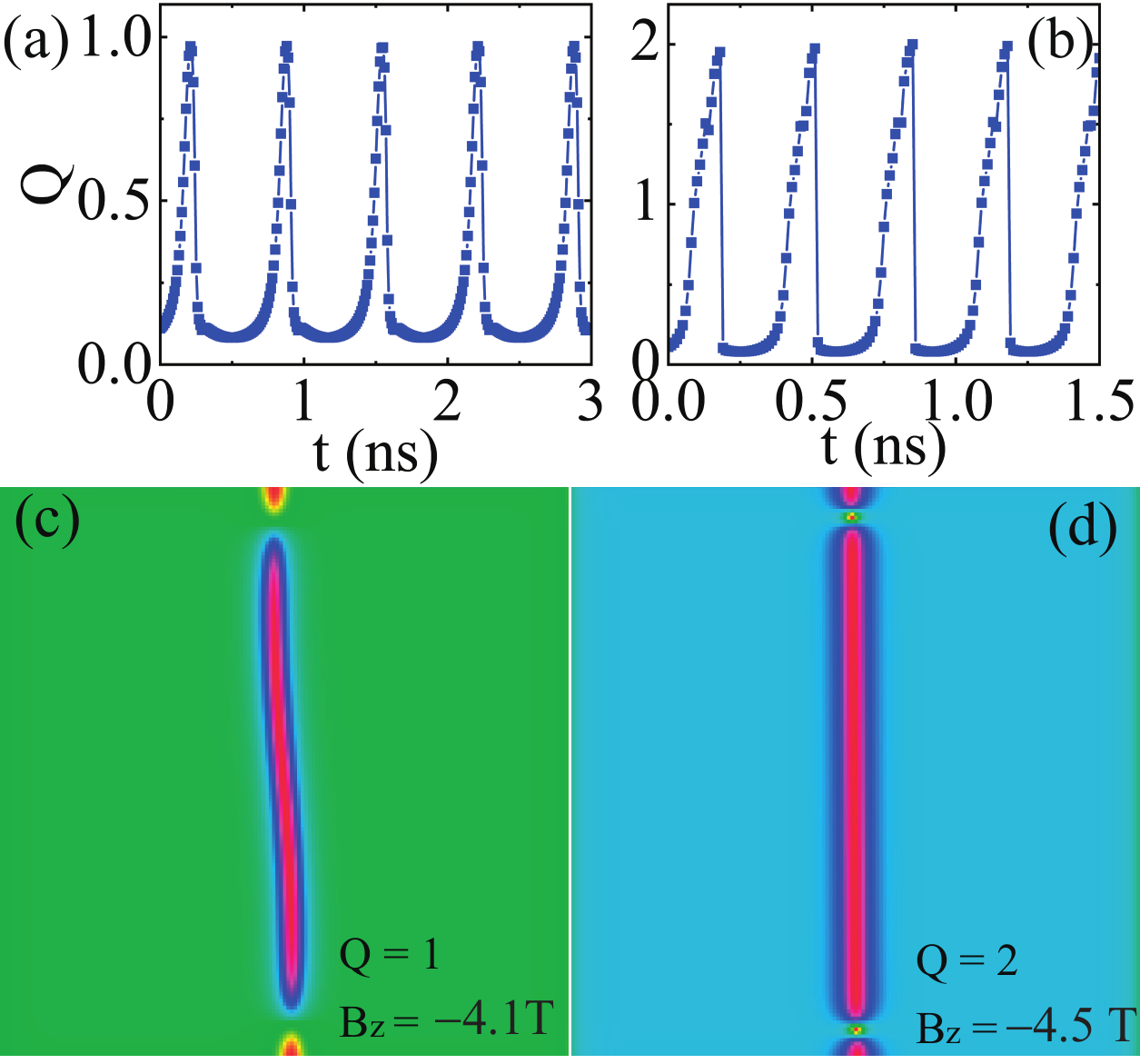}
\caption{The temporal periodic evolution of the topological charges for the alternating magnetic fields B with (a) $f=1.5$ GHz and $B_z=-4.1$ T, and (b) $f=3$ GHz and $B_z=-4.5$ T. The two non-trivial topological configurations (c) and (d) are generated periodically in (a) and (b), respectively.}
\label{fig11}
\end{figure}

The trivial and non-trivial topological configurations can switch periodically in the presence of an alternating magnetic field $B=B_z\sin(2\pi f t)$ with the frequency $f$ and time $t$, as illustrated in Fig. \ref{fig11} (a) for $Q=1$ and (b) for $Q=2$. $B_z$ is the strength of the perpendicular magnetic field. We have found that the dynamic configurations and the topological charges of the non-trivial topological states can be governed by the strength of the magnetic field [see Fig. \ref{fig11} (c) and (d)], and that the period of the transition can be tuned through the frequency of magnetic fields. The detailed switching processes are displayed in Videos 2 and 3 of the Supplementary Material at $f=1.5$ GHz and $B_z$=$-$4.1 T, and $f=3.0$ GHz and $B_z$=$-$4.5 T, respectively \cite{supply}, respectively. It shows that the time-periodic nanoscale chiral topological magnetic structures are achieved and controlled by the frequency and intensity of magnetic fields and the size of magnetic nanostructures. Our extensive simulations indicate that the frequency of the magnetic field determines the transition period between the topological states, while the strength of the magnetic field governs the structure of the non-trivial topological states and the topological charge.

The ability to control and encode data based on complex spin configurations and topological charges of skyrmions holds promise for developing next-generation spin storage nanodevices by an alternating magnetic field \cite{qin2c03046}. Our study provides a possible approach to achieve high-density, energy-efficient information storage in magnetic nanostructures. The devices would leverage the stable and tunable nature of such novel skyrmions, allowing for the creation of highly compact and low-power data storage systems. By utilizing the alternating magnetic fields required to manipulate the trivial and non-trivial topological configurations as ``0'' and ``1'' codes, respectively, we can envision the realization of spin-based memory and logic devices that could be used to process and store digital information. We believe that the potential to control and manipulate these distinctive magnetic textures using alternating magnetic fields in nanostructures holds enormous potential for the development of spin-memory nanodevices. Finally, it should be noted that, given the current experimental conditions, particularly those involving strong GHz alternating magnetic fields, the realization of our theoretical findings poses significant challenges. Presently, there is a dearth of research in this regard. Nevertheless, we are committing to the exploration with the aspiration that, under the usual experimental conditions and device requirements, we will be able to achieve the transformation of different topological states within artificial nanostructures readily.

\section{Conclusion}
The magnetic field and size-dependent tilted magnetic skyrmion states are induced in magnetic nanostructures with IMA under magnetic fields, due to geometric nanoconfinement effects. A controlled chain of periodic skyrmion states can be realized and tuned by the steady magnetic fields and the size of the nanoribbons. Moreover, tuning the size of such nanoribbons allows for the achievement of the tunable topological charges and no-integer topological charges of magnetic skyrmions. An alternating magnetic field can be used to periodically switch between trivial and non-trivial topological textures and the period of the switching is tuned by the frequency of magnetic fields. The tunable dynamic magnetic skyrmions are obtained in magnetic nanostructures. In comparison to magnetic skyrmions in PMA nanostructures and 2D magnetic materials, these findings are very different. These theoretical studies can be applied to understand and explain recent experimental results. Our study not only demonstrates novel skyrmion generation mechanisms and offers new ways to control skyrmions in nanostructures with the IMA, but also provides fresh insights for the design of spin nanodevices.

\par
\textit{Acknowledgments} - This work is supported by the National Natural Science Foundation of China (Grant Nos. 12374079 and 11604380).

\nocite{*}

\begin{thebibliography}{60}%
     \makeatletter
     \providecommand \@ifxundefined [1]{%
      \@ifx{#1\undefined}
     }%
     \providecommand \@ifnum [1]{%
      \ifnum #1\expandafter \@firstoftwo
      \else \expandafter \@secondoftwo
      \fi
     }%
     \providecommand \@ifx [1]{%
      \ifx #1\expandafter \@firstoftwo
      \else \expandafter \@secondoftwo
      \fi
     }%
     \providecommand \natexlab [1]{#1}%
     \providecommand \enquote  [1]{``#1''}%
     \providecommand \bibnamefont  [1]{#1}%
     \providecommand \bibfnamefont [1]{#1}%
     \providecommand \citenamefont [1]{#1}%
     \providecommand \href@noop [0]{\@secondoftwo}%
     \providecommand \href [0]{\begingroup \@sanitize@url \@href}%
     \providecommand \@href[1]{\@@startlink{#1}\@@href}%
     \providecommand \@@href[1]{\endgroup#1\@@endlink}%
     \providecommand \@sanitize@url [0]{\catcode `\\12\catcode `\$12\catcode `\&12\catcode `\#12\catcode `\^12\catcode `\_12\catcode `\%12\relax}%
     \providecommand \@@startlink[1]{}%
     \providecommand \@@endlink[0]{}%
     \providecommand \url  [0]{\begingroup\@sanitize@url \@url }%
     \providecommand \@url [1]{\endgroup\@href {#1}{\urlprefix }}%
     \providecommand \urlprefix  [0]{URL }%
     \providecommand \Eprint [0]{\href }%
     \providecommand \doibase [0]{https://doi.org/}%
     \providecommand \selectlanguage [0]{\@gobble}%
     \providecommand \bibinfo  [0]{\@secondoftwo}%
     \providecommand \bibfield  [0]{\@secondoftwo}%
     \providecommand \translation [1]{[#1]}%
     \providecommand \BibitemOpen [0]{}%
     \providecommand \bibitemStop [0]{}%
     \providecommand \bibitemNoStop [0]{.\EOS\space}%
     \providecommand \EOS [0]{\spacefactor3000\relax}%
     \providecommand \BibitemShut  [1]{\csname bibitem#1\endcsname}%
     \let\auto@bib@innerbib\@empty
     \bibitem [{\citenamefont {Jin}\ \emph {et~al.}(2017)\citenamefont {Jin}, \citenamefont {Li}, \citenamefont {Kov\'acs}, \citenamefont {Caron}, \citenamefont {Zheng}, \citenamefont {Rybakov}, \citenamefont {Kiselev}, \citenamefont {Du}, \citenamefont {Bl$\ddot{u}$gel}, \citenamefont {Tian}, \citenamefont {Zhang}, \citenamefont {Farle},\ and\ \citenamefont {Dunin-Borkowski}}]{jin_control_2017}%
       \BibitemOpen
       \bibfield  {author} {\bibinfo {author} {\bibfnamefont {C.}~\bibnamefont {Jin}}, \bibinfo {author} {\bibfnamefont {Z.~A.}\ \bibnamefont {Li}}, \bibinfo {author} {\bibfnamefont {A.}~\bibnamefont {Kov\'acs}}, \bibinfo {author} {\bibfnamefont {J.}~\bibnamefont {Caron}}, \bibinfo {author} {\bibfnamefont {F.~S.}\ \bibnamefont {Zheng}}, \bibinfo {author} {\bibfnamefont {F.~N.}\ \bibnamefont {Rybakov}}, \bibinfo {author} {\bibfnamefont {N.~S.}\ \bibnamefont {Kiselev}}, \bibinfo {author} {\bibfnamefont {H.~F.}\ \bibnamefont {Du}}, \bibinfo {author} {\bibfnamefont {S.}~\bibnamefont {Bl$\ddot{u}$gel}}, \bibinfo {author} {\bibfnamefont {M.~L.}\ \bibnamefont {Tian}}, \bibinfo {author} {\bibfnamefont {Y.~H.}\ \bibnamefont {Zhang}}, \bibinfo {author} {\bibfnamefont {M.}~\bibnamefont {Farle}},\ and\ \bibinfo {author} {\bibfnamefont {R.~E.}\ \bibnamefont {Dunin-Borkowski}},\ }\bibfield  {title} {\bibinfo {title} {Control of morphology and formation of highly geometrically confined magnetic skyrmions},\ }\href
       {https://doi.org/10.1038/ncomms15569} {\bibfield  {journal} {\bibinfo  {journal} {Nat. Commun.}\ }\textbf {\bibinfo {volume} {8}},\ \bibinfo {pages} {15569} (\bibinfo {year} {2017})}\BibitemShut {NoStop}%
     \bibitem [{\citenamefont {Rohart}\ and\ \citenamefont {Thiaville}(2013)}]{rohart_skyrmion_2013}%
       \BibitemOpen
       \bibfield  {author} {\bibinfo {author} {\bibfnamefont {S.}~\bibnamefont {Rohart}}\ and\ \bibinfo {author} {\bibfnamefont {A.}~\bibnamefont {Thiaville}},\ }\bibfield  {title} {\bibinfo {title} {Skyrmion confinement in ultrathin film nanostructures in the presence of {Dzyaloshinskii}-{Moriya} interaction},\ }\href {https://doi.org/10.1103/PhysRevB.88.184422} {\bibfield  {journal} {\bibinfo  {journal} {Phys. Rev. B}\ }\textbf {\bibinfo {volume} {88}},\ \bibinfo {pages} {184422} (\bibinfo {year} {2013})}\BibitemShut {NoStop}%
     \bibitem [{\citenamefont {Hagemeister}\ \emph {et~al.}(2016)\citenamefont {Hagemeister}, \citenamefont {Iaia}, \citenamefont {Vedmedenko}, \citenamefont {Von~Bergmann}, \citenamefont {Kubetzka},\ and\ \citenamefont {Wiesendanger}}]{hagemeister_2016}%
       \BibitemOpen
       \bibfield  {author} {\bibinfo {author} {\bibfnamefont {J.}~\bibnamefont {Hagemeister}}, \bibinfo {author} {\bibfnamefont {D.}~\bibnamefont {Iaia}}, \bibinfo {author} {\bibfnamefont {E.~Y.}\ \bibnamefont {Vedmedenko}}, \bibinfo {author} {\bibfnamefont {K.}~\bibnamefont {Von~Bergmann}}, \bibinfo {author} {\bibfnamefont {A.}~\bibnamefont {Kubetzka}},\ and\ \bibinfo {author} {\bibfnamefont {R.}~\bibnamefont {Wiesendanger}},\ }\bibfield  {title} {\bibinfo {title} {Skyrmions at the {Edge}: {Confinement} {Effects} in {Fe}/{Ir}( 111 )},\ }\href {https://doi.org/10.1103/PhysRevLett.117.207202} {\bibfield  {journal} {\bibinfo  {journal} {Phys. Rev. Lett.}\ }\textbf {\bibinfo {volume} {117}},\ \bibinfo {pages} {207202} (\bibinfo {year} {2016})}\BibitemShut {NoStop}%
     \bibitem [{\citenamefont {Navau}\ \emph {et~al.}(2016)\citenamefont {Navau}, \citenamefont {Del-Valle},\ and\ \citenamefont {Sanchez}}]{navau_analytical_2016}%
       \BibitemOpen
       \bibfield  {author} {\bibinfo {author} {\bibfnamefont {C.}~\bibnamefont {Navau}}, \bibinfo {author} {\bibfnamefont {N.}~\bibnamefont {Del-Valle}},\ and\ \bibinfo {author} {\bibfnamefont {A.}~\bibnamefont {Sanchez}},\ }\bibfield  {title} {\bibinfo {title} {Analytical trajectories of skyrmions in confined geometries: {Skyrmionic} racetracks and nano-oscillators},\ }\href {https://doi.org/10.1103/PhysRevB.94.184104} {\bibfield  {journal} {\bibinfo  {journal} {Phys. Rev. B}\ }\textbf {\bibinfo {volume} {94}},\ \bibinfo {pages} {184104} (\bibinfo {year} {2016})}\BibitemShut {NoStop}%
     \bibitem [{\citenamefont {Ho}\ \emph {et~al.}(2019)\citenamefont {Ho}, \citenamefont {Tan}, \citenamefont {Goolaup}, \citenamefont {Oyarce}, \citenamefont {Raju}, \citenamefont {Huang}, \citenamefont {Soumyanarayanan},\ and\ \citenamefont {Panagopoulos}}]{tan024064}%
       \BibitemOpen
       \bibfield  {author} {\bibinfo {author} {\bibfnamefont {P.}~\bibnamefont {Ho}}, \bibinfo {author} {\bibfnamefont {A.~K.~C.}\ \bibnamefont {Tan}}, \bibinfo {author} {\bibfnamefont {S.}~\bibnamefont {Goolaup}}, \bibinfo {author} {\bibfnamefont {A.~L.~G.}\ \bibnamefont {Oyarce}}, \bibinfo {author} {\bibfnamefont {M.}~\bibnamefont {Raju}}, \bibinfo {author} {\bibfnamefont {L.~S.}\ \bibnamefont {Huang}}, \bibinfo {author} {\bibfnamefont {A.}~\bibnamefont {Soumyanarayanan}},\ and\ \bibinfo {author} {\bibfnamefont {C.}~\bibnamefont {Panagopoulos}},\ }\bibfield  {title} {\bibinfo {title} {Geometrically tailored skyrmions at zero magnetic field in multilayered nanostructures},\ }\href {https://doi.org/10.1103/PhysRevApplied.11.024064} {\bibfield  {journal} {\bibinfo  {journal} {Phys. Rev. Appl.}\ }\textbf {\bibinfo {volume} {11}},\ \bibinfo {pages} {024064} (\bibinfo {year} {2019})}\BibitemShut {NoStop}%
     \bibitem [{\citenamefont {Tambovtsev}\ \emph {et~al.}(2022)\citenamefont {Tambovtsev}, \citenamefont {Leonov}, \citenamefont {Lobanov}, \citenamefont {Kiselev},\ and\ \citenamefont {Uzdin}}]{tambovtsev_2022}%
       \BibitemOpen
       \bibfield  {author} {\bibinfo {author} {\bibfnamefont {I.~M.}\ \bibnamefont {Tambovtsev}}, \bibinfo {author} {\bibfnamefont {A.~O.}\ \bibnamefont {Leonov}}, \bibinfo {author} {\bibfnamefont {I.~S.}\ \bibnamefont {Lobanov}}, \bibinfo {author} {\bibfnamefont {A.~D.}\ \bibnamefont {Kiselev}},\ and\ \bibinfo {author} {\bibfnamefont {V.~M.}\ \bibnamefont {Uzdin}},\ }\bibfield  {title} {\bibinfo {title} {Topological structures in chiral media: {Effects} of confined geometry},\ }\href {https://doi.org/10.1103/PhysRevE.105.034701} {\bibfield  {journal} {\bibinfo  {journal} {Phys. Rev. E}\ }\textbf {\bibinfo {volume} {105}},\ \bibinfo {pages} {034701} (\bibinfo {year} {2022})}\BibitemShut {NoStop}%
     \bibitem [{\citenamefont {Charilaou}\ \emph {et~al.}(2018)\citenamefont {Charilaou}, \citenamefont {Braun},\ and\ \citenamefont {L$\ddot{o}$ffler}}]{monopole_2018}%
       \BibitemOpen
       \bibfield  {author} {\bibinfo {author} {\bibfnamefont {M.}~\bibnamefont {Charilaou}}, \bibinfo {author} {\bibfnamefont {H.~B.}\ \bibnamefont {Braun}},\ and\ \bibinfo {author} {\bibfnamefont {J.~F.}\ \bibnamefont {L$\ddot{o}$ffler}},\ }\bibfield  {title} {\bibinfo {title} {Monopole-{Induced} {Emergent} {Electric} {Fields} in {Ferromagnetic} {Nanowires}},\ }\href {https://doi.org/10.1103/PhysRevLett.121.097202} {\bibfield  {journal} {\bibinfo  {journal} {Phys. Rev. Lett.}\ }\textbf {\bibinfo {volume} {121}},\ \bibinfo {pages} {097202} (\bibinfo {year} {2018})}\BibitemShut {NoStop}%
     \bibitem [{\citenamefont {Leonov}\ \emph {et~al.}(2016)\citenamefont {Leonov}, \citenamefont {Togawa}, \citenamefont {Monchesky}, \citenamefont {Bogdanov}, \citenamefont {Kishine}, \citenamefont {Kousaka}, \citenamefont {Miyagawa}, \citenamefont {Koyama}, \citenamefont {Akimitsu}, \citenamefont {Koyama}, \citenamefont {Harada}, \citenamefont {Mori}, \citenamefont {McGrouther}, \citenamefont {Lamb}, \citenamefont {Krajnak}, \citenamefont {McVitie}, \citenamefont {Stamps},\ and\ \citenamefont {Inoue}}]{leonov087202}%
       \BibitemOpen
       \bibfield  {author} {\bibinfo {author} {\bibfnamefont {A.~O.}\ \bibnamefont {Leonov}}, \bibinfo {author} {\bibfnamefont {Y.}~\bibnamefont {Togawa}}, \bibinfo {author} {\bibfnamefont {T.~L.}\ \bibnamefont {Monchesky}}, \bibinfo {author} {\bibfnamefont {A.~N.}\ \bibnamefont {Bogdanov}}, \bibinfo {author} {\bibfnamefont {J.}~\bibnamefont {Kishine}}, \bibinfo {author} {\bibfnamefont {Y.}~\bibnamefont {Kousaka}}, \bibinfo {author} {\bibfnamefont {M.}~\bibnamefont {Miyagawa}}, \bibinfo {author} {\bibfnamefont {T.}~\bibnamefont {Koyama}}, \bibinfo {author} {\bibfnamefont {J.}~\bibnamefont {Akimitsu}}, \bibinfo {author} {\bibfnamefont {T.}~\bibnamefont {Koyama}}, \bibinfo {author} {\bibfnamefont {K.}~\bibnamefont {Harada}}, \bibinfo {author} {\bibfnamefont {S.}~\bibnamefont {Mori}}, \bibinfo {author} {\bibfnamefont {D.}~\bibnamefont {McGrouther}}, \bibinfo {author} {\bibfnamefont {R.}~\bibnamefont {Lamb}}, \bibinfo {author} {\bibfnamefont {M.}~\bibnamefont {Krajnak}}, \bibinfo {author} {\bibfnamefont
       {S.}~\bibnamefont {McVitie}}, \bibinfo {author} {\bibfnamefont {R.~L.}\ \bibnamefont {Stamps}},\ and\ \bibinfo {author} {\bibfnamefont {K.}~\bibnamefont {Inoue}},\ }\bibfield  {title} {\bibinfo {title} {Chiral surface twists and skyrmion stability in nanolayers of cubic helimagnets},\ }\href {https://doi.org/10.1103/PhysRevLett.117.087202} {\bibfield  {journal} {\bibinfo  {journal} {Phys. Rev. Lett.}\ }\textbf {\bibinfo {volume} {117}},\ \bibinfo {pages} {087202} (\bibinfo {year} {2016})}\BibitemShut {NoStop}%
     \bibitem [{\citenamefont {Zhang}\ \emph {et~al.}(2020)\citenamefont {Zhang}, \citenamefont {Burn}, \citenamefont {Jaouen}, \citenamefont {Chauleau}, \citenamefont {Haghighirad}, \citenamefont {Liu}, \citenamefont {Wang}, \citenamefont {Van Der~Laan},\ and\ \citenamefont {Hesjedal}}]{zhang_robust_2020}%
       \BibitemOpen
       \bibfield  {author} {\bibinfo {author} {\bibfnamefont {S.~L.}\ \bibnamefont {Zhang}}, \bibinfo {author} {\bibfnamefont {D.~M.}\ \bibnamefont {Burn}}, \bibinfo {author} {\bibfnamefont {N.}~\bibnamefont {Jaouen}}, \bibinfo {author} {\bibfnamefont {J.-Y.}\ \bibnamefont {Chauleau}}, \bibinfo {author} {\bibfnamefont {A.~A.}\ \bibnamefont {Haghighirad}}, \bibinfo {author} {\bibfnamefont {Y.~Z.}\ \bibnamefont {Liu}}, \bibinfo {author} {\bibfnamefont {W.~W.}\ \bibnamefont {Wang}}, \bibinfo {author} {\bibfnamefont {G.}~\bibnamefont {Van Der~Laan}},\ and\ \bibinfo {author} {\bibfnamefont {T.}~\bibnamefont {Hesjedal}},\ }\bibfield  {title} {\bibinfo {title} {Robust {Perpendicular} {Skyrmions} and {Their} {Surface} {Confinement}},\ }\href {https://doi.org/10.1021/acs.nanolett.9b05141} {\bibfield  {journal} {\bibinfo  {journal} {Nano Lett.}\ }\textbf {\bibinfo {volume} {20}},\ \bibinfo {pages} {1428} (\bibinfo {year} {2020})}\BibitemShut {NoStop}%
     \bibitem [{\citenamefont {Zhao}\ \emph {et~al.}(2016)\citenamefont {Zhao}, \citenamefont {Jin}, \citenamefont {Wang}, \citenamefont {Du}, \citenamefont {Zang}, \citenamefont {Tian}, \citenamefont {Che},\ and\ \citenamefont {Zhang}}]{zhao_direct_2016}%
       \BibitemOpen
       \bibfield  {author} {\bibinfo {author} {\bibfnamefont {X.~B.}\ \bibnamefont {Zhao}}, \bibinfo {author} {\bibfnamefont {C.~M.}\ \bibnamefont {Jin}}, \bibinfo {author} {\bibfnamefont {C.}~\bibnamefont {Wang}}, \bibinfo {author} {\bibfnamefont {H.~F.}\ \bibnamefont {Du}}, \bibinfo {author} {\bibfnamefont {J.~D.}\ \bibnamefont {Zang}}, \bibinfo {author} {\bibfnamefont {M.~L.}\ \bibnamefont {Tian}}, \bibinfo {author} {\bibfnamefont {R.~C.}\ \bibnamefont {Che}},\ and\ \bibinfo {author} {\bibfnamefont {Y.~H.}\ \bibnamefont {Zhang}},\ }\bibfield  {title} {\bibinfo {title} {Direct imaging of magnetic field-driven transitions of skyrmion cluster states in {FeGe} nanodisks},\ }\href {https://doi.org/10.1073/pnas.1600197113} {\bibfield  {journal} {\bibinfo  {journal} {Proc. Natl. Acad. Sci.}\ }\textbf {\bibinfo {volume} {113}},\ \bibinfo {pages} {4918} (\bibinfo {year} {2016})}\BibitemShut {NoStop}%
     \bibitem [{\citenamefont {Yu}\ \emph {et~al.}(2010)\citenamefont {Yu}, \citenamefont {Onose}, \citenamefont {Kanazawa}, \citenamefont {Park}, \citenamefont {Han}, \citenamefont {Matsui}, \citenamefont {Nagaosa},\ and\ \citenamefont {Tokura}}]{Yu2010ayi}%
       \BibitemOpen
       \bibfield  {author} {\bibinfo {author} {\bibfnamefont {X.~Z.}\ \bibnamefont {Yu}}, \bibinfo {author} {\bibfnamefont {Y.}~\bibnamefont {Onose}}, \bibinfo {author} {\bibfnamefont {N.}~\bibnamefont {Kanazawa}}, \bibinfo {author} {\bibfnamefont {J.~H.}\ \bibnamefont {Park}}, \bibinfo {author} {\bibfnamefont {J.~H.}\ \bibnamefont {Han}}, \bibinfo {author} {\bibfnamefont {Y.}~\bibnamefont {Matsui}}, \bibinfo {author} {\bibfnamefont {N.}~\bibnamefont {Nagaosa}},\ and\ \bibinfo {author} {\bibfnamefont {Y.}~\bibnamefont {Tokura}},\ }\bibfield  {title} {\bibinfo {title} {Real-space observation of a two-dimensional skyrmion crystal},\ }\href {https://doi.org/10.1038/nature09124} {\bibfield  {journal} {\bibinfo  {journal} {Nature}\ }\textbf {\bibinfo {volume} {465}},\ \bibinfo {pages} {901} (\bibinfo {year} {2010})}\BibitemShut {NoStop}%
     \bibitem [{\citenamefont {Mulkers}\ \emph {et~al.}(2016)\citenamefont {Mulkers}, \citenamefont {Milo$\check{e}$vi$\acute{c}$},\ and\ \citenamefont {Van~Waeyenberge}}]{mulkers_cycloidal_2016}%
       \BibitemOpen
       \bibfield  {author} {\bibinfo {author} {\bibfnamefont {J.}~\bibnamefont {Mulkers}}, \bibinfo {author} {\bibfnamefont {M.~V.}\ \bibnamefont {Milo$\check{e}$vi$\acute{c}$}},\ and\ \bibinfo {author} {\bibfnamefont {B.}~\bibnamefont {Van~Waeyenberge}},\ }\bibfield  {title} {\bibinfo {title} {Cycloidal versus skyrmionic states in mesoscopic chiral magnets},\ }\href {https://doi.org/10.1103/PhysRevB.93.214405} {\bibfield  {journal} {\bibinfo  {journal} {Phys. Rev. B}\ }\textbf {\bibinfo {volume} {93}},\ \bibinfo {pages} {214405} (\bibinfo {year} {2016})}\BibitemShut {NoStop}%
     \bibitem [{\citenamefont {Mulkers}\ \emph {et~al.}(2017)\citenamefont {Mulkers}, \citenamefont {Van~Waeyenberge},\ and\ \citenamefont {Milo$\check{s}$evi\'c}}]{mulkers_2017}%
       \BibitemOpen
       \bibfield  {author} {\bibinfo {author} {\bibfnamefont {J.}~\bibnamefont {Mulkers}}, \bibinfo {author} {\bibfnamefont {B.}~\bibnamefont {Van~Waeyenberge}},\ and\ \bibinfo {author} {\bibfnamefont {M.~V.}\ \bibnamefont {Milo$\check{s}$evi\'c}},\ }\bibfield  {title} {\bibinfo {title} {Effects of spatially engineered {Dzyaloshinskii}-{Moriya} interaction in ferromagnetic films},\ }\href {https://doi.org/10.1103/PhysRevB.95.144401} {\bibfield  {journal} {\bibinfo  {journal} {Phys. Rev. B}\ }\textbf {\bibinfo {volume} {95}},\ \bibinfo {pages} {144401} (\bibinfo {year} {2017})}\BibitemShut {NoStop}%
     \bibitem [{\citenamefont {Bhattacharya}\ \emph {et~al.}(2020)\citenamefont {Bhattacharya}, \citenamefont {Razavi}, \citenamefont {Wu}, \citenamefont {Dai}, \citenamefont {Wang},\ and\ \citenamefont {Atulasimha}}]{Bhattacharya2020}%
       \BibitemOpen
       \bibfield  {author} {\bibinfo {author} {\bibfnamefont {D.}~\bibnamefont {Bhattacharya}}, \bibinfo {author} {\bibfnamefont {S.~A.}\ \bibnamefont {Razavi}}, \bibinfo {author} {\bibfnamefont {H.}~\bibnamefont {Wu}}, \bibinfo {author} {\bibfnamefont {B.}~\bibnamefont {Dai}}, \bibinfo {author} {\bibfnamefont {K.~L.}\ \bibnamefont {Wang}},\ and\ \bibinfo {author} {\bibfnamefont {J.}~\bibnamefont {Atulasimha}},\ }\bibfield  {title} {\bibinfo {title} {Creation and annihilation of non-volatile fixed magnetic skyrmions using voltage control of magnetic anisotropy},\ }\href {https://doi.org/10.1038/s41928-020-0432-x} {\bibfield  {journal} {\bibinfo  {journal} {Nature Electronics}\ }\textbf {\bibinfo {volume} {3}},\ \bibinfo {pages} {539} (\bibinfo {year} {2020})}\BibitemShut {NoStop}%
     \bibitem [{\citenamefont {Alshammari}\ \emph {et~al.}(2021)\citenamefont {Alshammari}, \citenamefont {Haltz}, \citenamefont {Alyami}, \citenamefont {Ali}, \citenamefont {Keatley}, \citenamefont {Marrows}, \citenamefont {Barker},\ and\ \citenamefont {Moore}}]{PRB224402}%
       \BibitemOpen
       \bibfield  {author} {\bibinfo {author} {\bibfnamefont {K.}~\bibnamefont {Alshammari}}, \bibinfo {author} {\bibfnamefont {E.}~\bibnamefont {Haltz}}, \bibinfo {author} {\bibfnamefont {M.}~\bibnamefont {Alyami}}, \bibinfo {author} {\bibfnamefont {M.}~\bibnamefont {Ali}}, \bibinfo {author} {\bibfnamefont {P.~S.}\ \bibnamefont {Keatley}}, \bibinfo {author} {\bibfnamefont {C.~H.}\ \bibnamefont {Marrows}}, \bibinfo {author} {\bibfnamefont {J.}~\bibnamefont {Barker}},\ and\ \bibinfo {author} {\bibfnamefont {T.~A.}\ \bibnamefont {Moore}},\ }\bibfield  {title} {\bibinfo {title} {Scaling of {D}zyaloshinskii-{M}oriya interaction with magnetization in {Pt/Co(Fe)B/Ir} multilayers},\ }\href {https://doi.org/10.1103/PhysRevB.104.224402} {\bibfield  {journal} {\bibinfo  {journal} {Phys. Rev. B}\ }\textbf {\bibinfo {volume} {104}},\ \bibinfo {pages} {224402} (\bibinfo {year} {2021})}\BibitemShut {NoStop}%
     \bibitem [{\citenamefont {Dong}\ \emph {et~al.}(2023)\citenamefont {Dong}, \citenamefont {Fu}, \citenamefont {Duan},\ and\ \citenamefont {Chang}}]{dong_tuning_2023}%
       \BibitemOpen
       \bibfield  {author} {\bibinfo {author} {\bibfnamefont {H.~M.}\ \bibnamefont {Dong}}, \bibinfo {author} {\bibfnamefont {P.~P.}\ \bibnamefont {Fu}}, \bibinfo {author} {\bibfnamefont {Y.~F.}\ \bibnamefont {Duan}},\ and\ \bibinfo {author} {\bibfnamefont {K.}~\bibnamefont {Chang}},\ }\bibfield  {title} {\bibinfo {title} {Tuning nano-skyrmions and nano-skyrmioniums in {Janus} magnets},\ }\href {https://doi.org/10.1039/D3NR02181E} {\bibfield  {journal} {\bibinfo  {journal} {Nanoscale}\ }\textbf {\bibinfo {volume} {15}},\ \bibinfo {pages} {15643} (\bibinfo {year} {2023})}\BibitemShut {NoStop}%
     \bibitem [{\citenamefont {Dong}\ \emph {et~al.}(2024)\citenamefont {Dong}, \citenamefont {Fu}, \citenamefont {Duan},\ and\ \citenamefont {Chang}}]{dongjap_2021}%
       \BibitemOpen
       \bibfield  {author} {\bibinfo {author} {\bibfnamefont {H.~M.}\ \bibnamefont {Dong}}, \bibinfo {author} {\bibfnamefont {P.~P.}\ \bibnamefont {Fu}}, \bibinfo {author} {\bibfnamefont {Y.~F.}\ \bibnamefont {Duan}},\ and\ \bibinfo {author} {\bibfnamefont {K.}~\bibnamefont {Chang}},\ }\bibfield  {title} {\bibinfo {title} {Ultrafast manipulations of nanoscale skyrmioniums},\ }\href {https://doi.org/10.1063/5.0227996} {\bibfield  {journal} {\bibinfo  {journal} {Journal of Applied Physics}\ }\textbf {\bibinfo {volume} {136}},\ \bibinfo {pages} {114301} (\bibinfo {year} {2024})}\BibitemShut {NoStop}%
     \bibitem [{\citenamefont {Yuan}\ \emph {et~al.}(2020)\citenamefont {Yuan}, \citenamefont {Yang}, \citenamefont {Cai}, \citenamefont {Wu}, \citenamefont {Chen}, \citenamefont {Yan},\ and\ \citenamefont {Shen}}]{intrinsic_2020}%
       \BibitemOpen
       \bibfield  {author} {\bibinfo {author} {\bibfnamefont {J.~R.}\ \bibnamefont {Yuan}}, \bibinfo {author} {\bibfnamefont {Y.~M.}\ \bibnamefont {Yang}}, \bibinfo {author} {\bibfnamefont {Y.~Q.}\ \bibnamefont {Cai}}, \bibinfo {author} {\bibfnamefont {Y.~H.}\ \bibnamefont {Wu}}, \bibinfo {author} {\bibfnamefont {Y.~P.}\ \bibnamefont {Chen}}, \bibinfo {author} {\bibfnamefont {X.~H.}\ \bibnamefont {Yan}},\ and\ \bibinfo {author} {\bibfnamefont {L.}~\bibnamefont {Shen}},\ }\bibfield  {title} {\bibinfo {title} {Intrinsic skyrmions in monolayer {Janus} magnets},\ }\href {https://doi.org/10.1103/PhysRevB.101.094420} {\bibfield  {journal} {\bibinfo  {journal} {Phys. Rev. B}\ }\textbf {\bibinfo {volume} {101}},\ \bibinfo {pages} {094420} (\bibinfo {year} {2020})}\BibitemShut {NoStop}%
     \bibitem [{\citenamefont {Im}\ \emph {et~al.}(2019)\citenamefont {Im}, \citenamefont {Han}, \citenamefont {Jung}, \citenamefont {Yu}, \citenamefont {Lee}, \citenamefont {Yoon}, \citenamefont {Chao}, \citenamefont {Fischer}, \citenamefont {Hong},\ and\ \citenamefont {Lee}}]{im_dynamics_2019}%
       \BibitemOpen
       \bibfield  {author} {\bibinfo {author} {\bibfnamefont {M.~Y.}\ \bibnamefont {Im}}, \bibinfo {author} {\bibfnamefont {H.~S.}\ \bibnamefont {Han}}, \bibinfo {author} {\bibfnamefont {M.~S.}\ \bibnamefont {Jung}}, \bibinfo {author} {\bibfnamefont {Y.~S.}\ \bibnamefont {Yu}}, \bibinfo {author} {\bibfnamefont {S.}~\bibnamefont {Lee}}, \bibinfo {author} {\bibfnamefont {S.}~\bibnamefont {Yoon}}, \bibinfo {author} {\bibfnamefont {W.~L.}\ \bibnamefont {Chao}}, \bibinfo {author} {\bibfnamefont {P.}~\bibnamefont {Fischer}}, \bibinfo {author} {\bibfnamefont {J.-I.}\ \bibnamefont {Hong}},\ and\ \bibinfo {author} {\bibfnamefont {K.-S.}\ \bibnamefont {Lee}},\ }\bibfield  {title} {\bibinfo {title} {Dynamics of the {Bloch} point in an asymmetric permalloy disk},\ }\href {https://doi.org/10.1038/s41467-019-08327-6} {\bibfield  {journal} {\bibinfo  {journal} {Nat. Commun.}\ }\textbf {\bibinfo {volume} {10}},\ \bibinfo {pages} {593} (\bibinfo {year} {2019})}\BibitemShut {NoStop}%
     \bibitem [{\citenamefont {Vousden}\ \emph {et~al.}(2016)\citenamefont {Vousden}, \citenamefont {Albert}, \citenamefont {Beg}, \citenamefont {Bisotti}, \citenamefont {Carey}, \citenamefont {Chernyshenko}, \citenamefont {Cort\'es-Ortu\~no}, \citenamefont {Wang}, \citenamefont {Hovorka}, \citenamefont {Marrows},\ and\ \citenamefont {Fangohr}}]{vousden_2016}%
       \BibitemOpen
       \bibfield  {author} {\bibinfo {author} {\bibfnamefont {M.}~\bibnamefont {Vousden}}, \bibinfo {author} {\bibfnamefont {M.}~\bibnamefont {Albert}}, \bibinfo {author} {\bibfnamefont {M.}~\bibnamefont {Beg}}, \bibinfo {author} {\bibfnamefont {M.-A.}\ \bibnamefont {Bisotti}}, \bibinfo {author} {\bibfnamefont {R.}~\bibnamefont {Carey}}, \bibinfo {author} {\bibfnamefont {D.}~\bibnamefont {Chernyshenko}}, \bibinfo {author} {\bibfnamefont {D.}~\bibnamefont {Cort\'es-Ortu\~no}}, \bibinfo {author} {\bibfnamefont {W.~W.}\ \bibnamefont {Wang}}, \bibinfo {author} {\bibfnamefont {O.}~\bibnamefont {Hovorka}}, \bibinfo {author} {\bibfnamefont {C.~H.}\ \bibnamefont {Marrows}},\ and\ \bibinfo {author} {\bibfnamefont {H.}~\bibnamefont {Fangohr}},\ }\bibfield  {title} {\bibinfo {title} {Skyrmions in thin films with easy-plane magnetocrystalline anisotropy},\ }\href {https://doi.org/10.1063/1.4945262} {\bibfield  {journal} {\bibinfo  {journal} {Applied Physics Letters}\ }\textbf {\bibinfo {volume} {108}},\ \bibinfo {pages}
       {132406} (\bibinfo {year} {2016})}\BibitemShut {NoStop}%
     \bibitem [{\citenamefont {Wilson}\ \emph {et~al.}(2014)\citenamefont {Wilson}, \citenamefont {Butenko}, \citenamefont {Bogdanov},\ and\ \citenamefont {Monchesky}}]{Wilson2014}%
       \BibitemOpen
       \bibfield  {author} {\bibinfo {author} {\bibfnamefont {M.~N.}\ \bibnamefont {Wilson}}, \bibinfo {author} {\bibfnamefont {A.~B.}\ \bibnamefont {Butenko}}, \bibinfo {author} {\bibfnamefont {A.~N.}\ \bibnamefont {Bogdanov}},\ and\ \bibinfo {author} {\bibfnamefont {T.~L.}\ \bibnamefont {Monchesky}},\ }\bibfield  {title} {\bibinfo {title} {Chiral skyrmions in cubic helimagnet films: The role of uniaxial anisotropy},\ }\href {https://doi.org/10.1103/PhysRevB.89.094411} {\bibfield  {journal} {\bibinfo  {journal} {Phys. Rev. B}\ }\textbf {\bibinfo {volume} {89}},\ \bibinfo {pages} {094411} (\bibinfo {year} {2014})}\BibitemShut {NoStop}%
     \bibitem [{\citenamefont {Guan}\ and\ \citenamefont {Ni}(2020{\natexlab{a}})}]{guan_2020}%
       \BibitemOpen
       \bibfield  {author} {\bibinfo {author} {\bibfnamefont {Z.~Y.}\ \bibnamefont {Guan}}\ and\ \bibinfo {author} {\bibfnamefont {S.}~\bibnamefont {Ni}},\ }\bibfield  {title} {\bibinfo {title} {Predicted {2D} ferromagnetic {Janus} {VSeTe} monolayer with high {Curie} temperature, large valley polarization and magnetic crystal anisotropy},\ }\href {https://doi.org/10.1039/D0NR04837B} {\bibfield  {journal} {\bibinfo  {journal} {Nanoscale}\ }\textbf {\bibinfo {volume} {12}},\ \bibinfo {pages} {22735} (\bibinfo {year} {2020}{\natexlab{a}})}\BibitemShut {NoStop}%
     \bibitem [{\citenamefont {Guan}\ and\ \citenamefont {Ni}(2020{\natexlab{b}})}]{guan13988}%
       \BibitemOpen
       \bibfield  {author} {\bibinfo {author} {\bibfnamefont {Z.~Y.}\ \bibnamefont {Guan}}\ and\ \bibinfo {author} {\bibfnamefont {S.}~\bibnamefont {Ni}},\ }\bibfield  {title} {\bibinfo {title} {{Strain-Controllable High Curie Temperature, Large Valley Polarization, and Magnetic Crystal Anisotropy in a 2D Ferromagnetic Janus VSeTe Monolayer}},\ }\href {https://doi.org/10.1021/acsami.0c13988} {\bibfield  {journal} {\bibinfo  {journal} {ACS Applied Materials \& Interfaces}\ }\textbf {\bibinfo {volume} {12}},\ \bibinfo {pages} {53067} (\bibinfo {year} {2020}{\natexlab{b}})}\BibitemShut {NoStop}%
     \bibitem [{\citenamefont {Smaili}\ \emph {et~al.}(2021)\citenamefont {Smaili}, \citenamefont {Laref}, \citenamefont {Garcia}, \citenamefont {Schwingenschl$\ddot{o}$gl}, \citenamefont {Roche},\ and\ \citenamefont {Manchon}}]{smaili_2021}%
       \BibitemOpen
       \bibfield  {author} {\bibinfo {author} {\bibfnamefont {I.}~\bibnamefont {Smaili}}, \bibinfo {author} {\bibfnamefont {S.}~\bibnamefont {Laref}}, \bibinfo {author} {\bibfnamefont {J.~H.}\ \bibnamefont {Garcia}}, \bibinfo {author} {\bibfnamefont {U.}~\bibnamefont {Schwingenschl$\ddot{o}$gl}}, \bibinfo {author} {\bibfnamefont {S.}~\bibnamefont {Roche}},\ and\ \bibinfo {author} {\bibfnamefont {A.}~\bibnamefont {Manchon}},\ }\bibfield  {title} {\bibinfo {title} {Janus monolayers of magnetic transition metal dichalcogenides as an all-in-one platform for spin-orbit torque},\ }\href {https://doi.org/10.1103/PhysRevB.104.104415} {\bibfield  {journal} {\bibinfo  {journal} {Phys. Rev. B}\ }\textbf {\bibinfo {volume} {104}},\ \bibinfo {pages} {104415} (\bibinfo {year} {2021})}\BibitemShut {NoStop}%
     \bibitem [{\citenamefont {Banerjee}\ \emph {et~al.}(2014)\citenamefont {Banerjee}, \citenamefont {Rowland}, \citenamefont {Erten},\ and\ \citenamefont {Randeria}}]{banerjee2014}%
       \BibitemOpen
       \bibfield  {author} {\bibinfo {author} {\bibfnamefont {S.}~\bibnamefont {Banerjee}}, \bibinfo {author} {\bibfnamefont {J.}~\bibnamefont {Rowland}}, \bibinfo {author} {\bibfnamefont {O.}~\bibnamefont {Erten}},\ and\ \bibinfo {author} {\bibfnamefont {M.}~\bibnamefont {Randeria}},\ }\bibfield  {title} {\bibinfo {title} {Enhanced stability of skyrmions in two-dimensional chiral magnets with rashba spin-orbit coupling},\ }\href {https://doi.org/10.1103/PhysRevX.4.031045} {\bibfield  {journal} {\bibinfo  {journal} {Phys. Rev. X}\ }\textbf {\bibinfo {volume} {4}},\ \bibinfo {pages} {031045} (\bibinfo {year} {2014})}\BibitemShut {NoStop}%
     \bibitem [{\citenamefont {Zhou}\ \emph {et~al.}(2024)\citenamefont {Zhou}, \citenamefont {Xu},\ and\ \citenamefont {Wu}}]{zhou_giant_2024}%
       \BibitemOpen
       \bibfield  {author} {\bibinfo {author} {\bibfnamefont {C.}~\bibnamefont {Zhou}}, \bibinfo {author} {\bibfnamefont {J.}~\bibnamefont {Xu}},\ and\ \bibinfo {author} {\bibfnamefont {Y.~Z.}\ \bibnamefont {Wu}},\ }\bibfield  {title} {\bibinfo {title} {Giant interfacial in-plane magnetic anisotropy in {Co}/{Pt} bilayers grown on {MgO}(110) substrates},\ }\href {https://doi.org/10.1103/PhysRevMaterials.8.024408} {\bibfield  {journal} {\bibinfo  {journal} {Phys. Rev. Materials}\ }\textbf {\bibinfo {volume} {8}},\ \bibinfo {pages} {024408} (\bibinfo {year} {2024})}\BibitemShut {NoStop}%
     \bibitem [{\citenamefont {Chen}\ \emph {et~al.}(2016)\citenamefont {Chen}, \citenamefont {Zhang},\ and\ \citenamefont {Liu}}]{chen_exotic_2016}%
       \BibitemOpen
       \bibfield  {author} {\bibinfo {author} {\bibfnamefont {J.~P.}\ \bibnamefont {Chen}}, \bibinfo {author} {\bibfnamefont {D.~W.}\ \bibnamefont {Zhang}},\ and\ \bibinfo {author} {\bibfnamefont {J.~M.}\ \bibnamefont {Liu}},\ }\bibfield  {title} {\bibinfo {title} {Exotic skyrmion crystals in chiral magnets with compass anisotropy},\ }\href {https://doi.org/10.1038/srep29126} {\bibfield  {journal} {\bibinfo  {journal} {Sci. Rep.}\ }\textbf {\bibinfo {volume} {6}},\ \bibinfo {pages} {29126} (\bibinfo {year} {2016})}\BibitemShut {NoStop}%
     \bibitem [{\citenamefont {Hayami}(2021)}]{hayami224418}%
       \BibitemOpen
       \bibfield  {author} {\bibinfo {author} {\bibfnamefont {S.}~\bibnamefont {Hayami}},\ }\bibfield  {title} {\bibinfo {title} {In-plane magnetic field-induced skyrmion crystal in frustrated magnets with easy-plane anisotropy},\ }\href {https://doi.org/10.1103/PhysRevB.103.224418} {\bibfield  {journal} {\bibinfo  {journal} {Phys. Rev. B}\ }\textbf {\bibinfo {volume} {103}},\ \bibinfo {pages} {224418} (\bibinfo {year} {2021})}\BibitemShut {NoStop}%
     \bibitem [{\citenamefont {Cheng}\ \emph {et~al.}(2021)\citenamefont {Cheng}, \citenamefont {Yan}, \citenamefont {Dong}, \citenamefont {Liu}, \citenamefont {Xia}, \citenamefont {Li},\ and\ \citenamefont {Han}}]{chen174409}%
       \BibitemOpen
       \bibfield  {author} {\bibinfo {author} {\bibfnamefont {C.}~\bibnamefont {Cheng}}, \bibinfo {author} {\bibfnamefont {Z.~R.}\ \bibnamefont {Yan}}, \bibinfo {author} {\bibfnamefont {J.}~\bibnamefont {Dong}}, \bibinfo {author} {\bibfnamefont {Y.~Z.}\ \bibnamefont {Liu}}, \bibinfo {author} {\bibfnamefont {Z.~C.}\ \bibnamefont {Xia}}, \bibinfo {author} {\bibfnamefont {L.}~\bibnamefont {Li}},\ and\ \bibinfo {author} {\bibfnamefont {X.~F.}\ \bibnamefont {Han}},\ }\bibfield  {title} {\bibinfo {title} {Elliptical skyrmion moving along a track without transverse speed},\ }\href {https://doi.org/10.1103/PhysRevB.104.174409} {\bibfield  {journal} {\bibinfo  {journal} {Phys. Rev. B}\ }\textbf {\bibinfo {volume} {104}},\ \bibinfo {pages} {174409} (\bibinfo {year} {2021})}\BibitemShut {NoStop}%
     \bibitem [{\citenamefont {Li}\ \emph {et~al.}(2024)\citenamefont {Li}, \citenamefont {Haldar}, \citenamefont {Kollwitz}, \citenamefont {Schrautzer}, \citenamefont {Goerzen},\ and\ \citenamefont {Heinze}}]{lil220404}%
       \BibitemOpen
       \bibfield  {author} {\bibinfo {author} {\bibfnamefont {D.~Z.}\ \bibnamefont {Li}}, \bibinfo {author} {\bibfnamefont {S.}~\bibnamefont {Haldar}}, \bibinfo {author} {\bibfnamefont {L.}~\bibnamefont {Kollwitz}}, \bibinfo {author} {\bibfnamefont {H.}~\bibnamefont {Schrautzer}}, \bibinfo {author} {\bibfnamefont {M.~A.}\ \bibnamefont {Goerzen}},\ and\ \bibinfo {author} {\bibfnamefont {S.}~\bibnamefont {Heinze}},\ }\bibfield  {title} {\bibinfo {title} {Prediction of stable nanoscale skyrmions in monolayer {Fe}$_5${GeTe}$_2$},\ }\href {https://doi.org/10.1103/PhysRevB.109.L220404} {\bibfield  {journal} {\bibinfo  {journal} {Phys. Rev. B}\ }\textbf {\bibinfo {volume} {109}},\ \bibinfo {pages} {L220404} (\bibinfo {year} {2024})}\BibitemShut {NoStop}%
     \bibitem [{\citenamefont {Flacke}\ \emph {et~al.}(2021)\citenamefont {Flacke}, \citenamefont {Ahrens}, \citenamefont {Mendisch}, \citenamefont {K\"orber}, \citenamefont {B\"ottcher}, \citenamefont {Meidinger}, \citenamefont {Yaqoob}, \citenamefont {M\"uller}, \citenamefont {Liensberger}, \citenamefont {K\'akay}, \citenamefont {Becherer}, \citenamefont {Pirro}, \citenamefont {Althammer}, \citenamefont {Gepr\"ags}, \citenamefont {Huebl}, \citenamefont {Gross},\ and\ \citenamefont {Weiler}}]{flacke_robust_2021}%
       \BibitemOpen
       \bibfield  {author} {\bibinfo {author} {\bibfnamefont {L.}~\bibnamefont {Flacke}}, \bibinfo {author} {\bibfnamefont {V.}~\bibnamefont {Ahrens}}, \bibinfo {author} {\bibfnamefont {S.}~\bibnamefont {Mendisch}}, \bibinfo {author} {\bibfnamefont {L.}~\bibnamefont {K\"orber}}, \bibinfo {author} {\bibfnamefont {T.}~\bibnamefont {B\"ottcher}}, \bibinfo {author} {\bibfnamefont {E.}~\bibnamefont {Meidinger}}, \bibinfo {author} {\bibfnamefont {M.}~\bibnamefont {Yaqoob}}, \bibinfo {author} {\bibfnamefont {M.}~\bibnamefont {M\"uller}}, \bibinfo {author} {\bibfnamefont {L.}~\bibnamefont {Liensberger}}, \bibinfo {author} {\bibfnamefont {A.}~\bibnamefont {K\'akay}}, \bibinfo {author} {\bibfnamefont {M.}~\bibnamefont {Becherer}}, \bibinfo {author} {\bibfnamefont {P.}~\bibnamefont {Pirro}}, \bibinfo {author} {\bibfnamefont {M.}~\bibnamefont {Althammer}}, \bibinfo {author} {\bibfnamefont {S.}~\bibnamefont {Gepr\"ags}}, \bibinfo {author} {\bibfnamefont {H.}~\bibnamefont {Huebl}}, \bibinfo {author} {\bibfnamefont
       {R.}~\bibnamefont {Gross}},\ and\ \bibinfo {author} {\bibfnamefont {M.}~\bibnamefont {Weiler}},\ }\bibfield  {title} {\bibinfo {title} {Robust formation of nanoscale magnetic skyrmions in easy-plane anisotropy thin film multilayers with low damping},\ }\href {https://doi.org/10.1103/PhysRevB.104.L100417} {\bibfield  {journal} {\bibinfo  {journal} {Phys. Rev. B}\ }\textbf {\bibinfo {volume} {104}},\ \bibinfo {pages} {L100417} (\bibinfo {year} {2021})}\BibitemShut {NoStop}%
     \bibitem [{\citenamefont {Montoya}\ \emph {et~al.}(2023)\citenamefont {Montoya}, \citenamefont {Khan}, \citenamefont {Safranski}, \citenamefont {Smith},\ and\ \citenamefont {Krivorotov}}]{montoya_2023}%
       \BibitemOpen
       \bibfield  {author} {\bibinfo {author} {\bibfnamefont {E.~A.}\ \bibnamefont {Montoya}}, \bibinfo {author} {\bibfnamefont {A.}~\bibnamefont {Khan}}, \bibinfo {author} {\bibfnamefont {C.}~\bibnamefont {Safranski}}, \bibinfo {author} {\bibfnamefont {A.}~\bibnamefont {Smith}},\ and\ \bibinfo {author} {\bibfnamefont {I.~N.}\ \bibnamefont {Krivorotov}},\ }\bibfield  {title} {\bibinfo {title} {Easy-plane spin {Hall} oscillator},\ }\href {https://doi.org/10.1038/s42005-023-01298-7} {\bibfield  {journal} {\bibinfo  {journal} {Commun. Phys.}\ }\textbf {\bibinfo {volume} {6}},\ \bibinfo {pages} {184} (\bibinfo {year} {2023})}\BibitemShut {NoStop}%
     \bibitem [{\citenamefont {Luo}\ \emph {et~al.}(2024)\citenamefont {Luo}, \citenamefont {Yu}, \citenamefont {Li}, \citenamefont {Qiu}, \citenamefont {Wang}, \citenamefont {Zhu},\ and\ \citenamefont {Zhou}}]{luo_ferrom_2024}%
       \BibitemOpen
       \bibfield  {author} {\bibinfo {author} {\bibfnamefont {D.~Y.}\ \bibnamefont {Luo}}, \bibinfo {author} {\bibfnamefont {G.~L.}\ \bibnamefont {Yu}}, \bibinfo {author} {\bibfnamefont {Y.}~\bibnamefont {Li}}, \bibinfo {author} {\bibfnamefont {Y.}~\bibnamefont {Qiu}}, \bibinfo {author} {\bibfnamefont {J.~W.}\ \bibnamefont {Wang}}, \bibinfo {author} {\bibfnamefont {M.~M.}\ \bibnamefont {Zhu}},\ and\ \bibinfo {author} {\bibfnamefont {H.~M.}\ \bibnamefont {Zhou}},\ }\bibfield  {title} {\bibinfo {title} {A ferromagnetic skyrmion-based spin-torque nano-oscillator with modified edge magnetization},\ }\href {https://doi.org/10.1088/1361-6463/ad2e4e} {\bibfield  {journal} {\bibinfo  {journal} {J. Phys. D: Appl. Phys.}\ }\textbf {\bibinfo {volume} {57}},\ \bibinfo {pages} {235001} (\bibinfo {year} {2024})}\BibitemShut {NoStop}%
     \bibitem [{\citenamefont {Vigo-Cotrina}\ and\ \citenamefont {Guimar$\tilde{a}$es}(2020)}]{vigo2020}%
       \BibitemOpen
       \bibfield  {author} {\bibinfo {author} {\bibfnamefont {H.}~\bibnamefont {Vigo-Cotrina}}\ and\ \bibinfo {author} {\bibfnamefont {A.~P.}\ \bibnamefont {Guimar$\tilde{a}$es}},\ }\bibfield  {title} {\bibinfo {title} {Creating skyrmions and skyrmioniums using oscillating perpendicular magnetic fields},\ }\href {https://doi.org/10.1016/j.jmmm.2020.166848} {\bibfield  {journal} {\bibinfo  {journal} {Journal of Magnetism and Magnetic Materials}\ }\textbf {\bibinfo {volume} {507}},\ \bibinfo {pages} {166848} (\bibinfo {year} {2020})}\BibitemShut {NoStop}%
     \bibitem [{\citenamefont {Moon}\ \emph {et~al.}(2016)\citenamefont {Moon}, \citenamefont {Kim}, \citenamefont {Je}, \citenamefont {Chun}, \citenamefont {Kim}, \citenamefont {Qiu}, \citenamefont {Choe},\ and\ \citenamefont {Hwang}}]{moon_skyrmion_2016}%
       \BibitemOpen
       \bibfield  {author} {\bibinfo {author} {\bibfnamefont {K.~W.}\ \bibnamefont {Moon}}, \bibinfo {author} {\bibfnamefont {D.~H.}\ \bibnamefont {Kim}}, \bibinfo {author} {\bibfnamefont {S.~G.}\ \bibnamefont {Je}}, \bibinfo {author} {\bibfnamefont {B.~S.}\ \bibnamefont {Chun}}, \bibinfo {author} {\bibfnamefont {W.}~\bibnamefont {Kim}}, \bibinfo {author} {\bibfnamefont {Z.~Q.}\ \bibnamefont {Qiu}}, \bibinfo {author} {\bibfnamefont {S.~B.}\ \bibnamefont {Choe}},\ and\ \bibinfo {author} {\bibfnamefont {C.~Y.}\ \bibnamefont {Hwang}},\ }\bibfield  {title} {\bibinfo {title} {Skyrmion motion driven by oscillating magnetic field},\ }\href {https://doi.org/10.1038/srep20360} {\bibfield  {journal} {\bibinfo  {journal} {Sci. Rep.}\ }\textbf {\bibinfo {volume} {6}},\ \bibinfo {pages} {20360} (\bibinfo {year} {2016})}\BibitemShut {NoStop}%
     \bibitem [{\citenamefont {Goto}\ \emph {et~al.}(2021)\citenamefont {Goto}, \citenamefont {Nomura},\ and\ \citenamefont {Suzuki}}]{stochastic_2021}%
       \BibitemOpen
       \bibfield  {author} {\bibinfo {author} {\bibfnamefont {M.}~\bibnamefont {Goto}}, \bibinfo {author} {\bibfnamefont {H.}~\bibnamefont {Nomura}},\ and\ \bibinfo {author} {\bibfnamefont {Y.}~\bibnamefont {Suzuki}},\ }\bibfield  {title} {\bibinfo {title} {Stochastic skyrmion dynamics under alternating magnetic fields},\ }\href {https://doi.org/10.1016/j.jmmm.2021.167974} {\bibfield  {journal} {\bibinfo  {journal} {Journal of Magnetism and Magnetic Materials}\ }\textbf {\bibinfo {volume} {536}},\ \bibinfo {pages} {167974} (\bibinfo {year} {2021})}\BibitemShut {NoStop}%
     \bibitem [{\citenamefont {Yang}\ \emph {et~al.}(2021)\citenamefont {Yang}, \citenamefont {Park}, \citenamefont {Park}, \citenamefont {Abert}, \citenamefont {Suess},\ and\ \citenamefont {Kim}}]{yan134427}%
       \BibitemOpen
       \bibfield  {author} {\bibinfo {author} {\bibfnamefont {J.}~\bibnamefont {Yang}}, \bibinfo {author} {\bibfnamefont {H.~K.}\ \bibnamefont {Park}}, \bibinfo {author} {\bibfnamefont {G.}~\bibnamefont {Park}}, \bibinfo {author} {\bibfnamefont {C.}~\bibnamefont {Abert}}, \bibinfo {author} {\bibfnamefont {D.}~\bibnamefont {Suess}},\ and\ \bibinfo {author} {\bibfnamefont {S.~K.}\ \bibnamefont {Kim}},\ }\bibfield  {title} {\bibinfo {title} {Robust formation of skyrmion and skyrmionium in magnetic hemispherical shells and their dynamic switching},\ }\href {https://doi.org/10.1103/PhysRevB.104.134427} {\bibfield  {journal} {\bibinfo  {journal} {Phys. Rev. B}\ }\textbf {\bibinfo {volume} {104}},\ \bibinfo {pages} {134427} (\bibinfo {year} {2021})}\BibitemShut {NoStop}%
     \bibitem [{\citenamefont {Shen}\ \emph {et~al.}(2020)\citenamefont {Shen}, \citenamefont {Li}, \citenamefont {Xia}, \citenamefont {Qiu}, \citenamefont {Zhang}, \citenamefont {Tretiakov}, \citenamefont {Ezawa},\ and\ \citenamefont {Zhou}}]{shen_dynamics_2020}%
       \BibitemOpen
       \bibfield  {author} {\bibinfo {author} {\bibfnamefont {L.~C.}\ \bibnamefont {Shen}}, \bibinfo {author} {\bibfnamefont {X.~G.}\ \bibnamefont {Li}}, \bibinfo {author} {\bibfnamefont {J.}~\bibnamefont {Xia}}, \bibinfo {author} {\bibfnamefont {L.}~\bibnamefont {Qiu}}, \bibinfo {author} {\bibfnamefont {X.~C.}\ \bibnamefont {Zhang}}, \bibinfo {author} {\bibfnamefont {O.~A.}\ \bibnamefont {Tretiakov}}, \bibinfo {author} {\bibfnamefont {M.}~\bibnamefont {Ezawa}},\ and\ \bibinfo {author} {\bibfnamefont {Y.}~\bibnamefont {Zhou}},\ }\bibfield  {title} {\bibinfo {title} {Dynamics of ferromagnetic bimerons driven by spin currents and magnetic fields},\ }\href {https://doi.org/10.1103/PhysRevB.102.104427} {\bibfield  {journal} {\bibinfo  {journal} {Phys. Rev. B}\ }\textbf {\bibinfo {volume} {102}},\ \bibinfo {pages} {104427} (\bibinfo {year} {2020})}\BibitemShut {NoStop}%
     \bibitem [{\citenamefont {Park}\ and\ \citenamefont {Kim}(2023)}]{park_emergence_2023}%
       \BibitemOpen
       \bibfield  {author} {\bibinfo {author} {\bibfnamefont {G.}~\bibnamefont {Park}}\ and\ \bibinfo {author} {\bibfnamefont {S.~K.}\ \bibnamefont {Kim}},\ }\bibfield  {title} {\bibinfo {title} {Emergence of chaos in magnetic-field-driven skyrmions},\ }\href {https://doi.org/10.1103/PhysRevB.108.174441} {\bibfield  {journal} {\bibinfo  {journal} {Phys. Rev. B}\ }\textbf {\bibinfo {volume} {108}},\ \bibinfo {pages} {174441} (\bibinfo {year} {2023})}\BibitemShut {NoStop}%
     \bibitem [{\citenamefont {Evans}\ \emph {et~al.}(2014)\citenamefont {Evans}, \citenamefont {Fan}, \citenamefont {Chureemart}, \citenamefont {Ostler}, \citenamefont {Ellis},\ and\ \citenamefont {Chantrell}}]{evans_atomistic_2014}%
       \BibitemOpen
       \bibfield  {author} {\bibinfo {author} {\bibfnamefont {R.~F.~L.}\ \bibnamefont {Evans}}, \bibinfo {author} {\bibfnamefont {W.~J.}\ \bibnamefont {Fan}}, \bibinfo {author} {\bibfnamefont {P.}~\bibnamefont {Chureemart}}, \bibinfo {author} {\bibfnamefont {T.~A.}\ \bibnamefont {Ostler}}, \bibinfo {author} {\bibfnamefont {M.~O.~A.}\ \bibnamefont {Ellis}},\ and\ \bibinfo {author} {\bibfnamefont {R.~W.}\ \bibnamefont {Chantrell}},\ }\bibfield  {title} {\bibinfo {title} {Atomistic spin model simulations of magnetic nanomaterials},\ }\href {https://doi.org/10.1088/0953-8984/26/10/103202} {\bibfield  {journal} {\bibinfo  {journal} {J. Phys.: Condens. Matter}\ }\textbf {\bibinfo {volume} {26}},\ \bibinfo {pages} {103202} (\bibinfo {year} {2014})}\BibitemShut {NoStop}%
     \bibitem [{\citenamefont {Tai}\ \emph {et~al.}(2024)\citenamefont {Tai}, \citenamefont {Hess}, \citenamefont {Wu},\ and\ \citenamefont {Smalyukh}}]{tai_2024}%
       \BibitemOpen
       \bibfield  {author} {\bibinfo {author} {\bibfnamefont {J.~S.~B.}\ \bibnamefont {Tai}}, \bibinfo {author} {\bibfnamefont {A.~J.}\ \bibnamefont {Hess}}, \bibinfo {author} {\bibfnamefont {J.~S.}\ \bibnamefont {Wu}},\ and\ \bibinfo {author} {\bibfnamefont {I.~I.}\ \bibnamefont {Smalyukh}},\ }\bibfield  {title} {\bibinfo {title} {Field-controlled dynamics of skyrmions and monopoles},\ }\href {https://doi.org/10.1126/sciadv.adj9373} {\bibfield  {journal} {\bibinfo  {journal} {Sci. Adv.}\ }\textbf {\bibinfo {volume} {10}},\ \bibinfo {pages} {eadj9373} (\bibinfo {year} {2024})}\BibitemShut {NoStop}%
     \bibitem [{\citenamefont {Aranda}\ and\ \citenamefont {Guslienko}(2018)}]{ma11112238}%
       \BibitemOpen
       \bibfield  {author} {\bibinfo {author} {\bibfnamefont {A.~R.}\ \bibnamefont {Aranda}}\ and\ \bibinfo {author} {\bibfnamefont {K.~Y.}\ \bibnamefont {Guslienko}},\ }\bibfield  {title} {\bibinfo {title} {Single chiral skyrmions in ultrathin magnetic films},\ }\href {https://doi.org/10.3390/ma11112238} {\bibfield  {journal} {\bibinfo  {journal} {Materials}\ }\textbf {\bibinfo {volume} {11}},\ \bibinfo {pages} {2238} (\bibinfo {year} {2018})}\BibitemShut {NoStop}%
     \bibitem [{\citenamefont {Butenko}\ \emph {et~al.}(2010)\citenamefont {Butenko}, \citenamefont {Leonov}, \citenamefont {R{\"o}ler},\ and\ \citenamefont {Bogdanov}}]{butenko_2010}%
       \BibitemOpen
       \bibfield  {author} {\bibinfo {author} {\bibfnamefont {A.~B.}\ \bibnamefont {Butenko}}, \bibinfo {author} {\bibfnamefont {A.~A.}\ \bibnamefont {Leonov}}, \bibinfo {author} {\bibfnamefont {U.~K.}\ \bibnamefont {R{\"o}ler}},\ and\ \bibinfo {author} {\bibfnamefont {A.~N.}\ \bibnamefont {Bogdanov}},\ }\bibfield  {title} {\bibinfo {title} {Stabilization of skyrmion textures by uniaxial distortions in noncentrosymmetric cubic helimagnets},\ }\href {https://doi.org/10.1103/PhysRevB.82.052403} {\bibfield  {journal} {\bibinfo  {journal} {Phys. Rev. B}\ }\textbf {\bibinfo {volume} {82}},\ \bibinfo {pages} {052403} (\bibinfo {year} {2010})}\BibitemShut {NoStop}%
     \bibitem [{\citenamefont {Kobayashi}\ and\ \citenamefont {Nitta}(2013)}]{kobayashi_2013}%
       \BibitemOpen
       \bibfield  {author} {\bibinfo {author} {\bibfnamefont {M.}~\bibnamefont {Kobayashi}}\ and\ \bibinfo {author} {\bibfnamefont {M.}~\bibnamefont {Nitta}},\ }\bibfield  {title} {\bibinfo {title} {Sine-{Gordon} kinks on a domain wall ring},\ }\href {https://doi.org/10.1103/PhysRevD.87.085003} {\bibfield  {journal} {\bibinfo  {journal} {Phys. Rev. D}\ }\textbf {\bibinfo {volume} {87}},\ \bibinfo {pages} {085003} (\bibinfo {year} {2013})}\BibitemShut {NoStop}%
     \bibitem [{\citenamefont {Jiang}\ \emph {et~al.}(2024)\citenamefont {Jiang}, \citenamefont {Zhou}, \citenamefont {Zhang},\ and\ \citenamefont {Mochizuki}}]{jiang013229}%
       \BibitemOpen
       \bibfield  {author} {\bibinfo {author} {\bibfnamefont {A.}~\bibnamefont {Jiang}}, \bibinfo {author} {\bibfnamefont {Y.}~\bibnamefont {Zhou}}, \bibinfo {author} {\bibfnamefont {X.}~\bibnamefont {Zhang}},\ and\ \bibinfo {author} {\bibfnamefont {M.}~\bibnamefont {Mochizuki}},\ }\bibfield  {title} {\bibinfo {title} {Transformation of a skyrmionium to a skyrmion through the thermal annihilation of the inner skyrmion},\ }\href {https://doi.org/10.1103/PhysRevResearch.6.013229} {\bibfield  {journal} {\bibinfo  {journal} {Phys. Rev. Res.}\ }\textbf {\bibinfo {volume} {6}},\ \bibinfo {pages} {013229} (\bibinfo {year} {2024})}\BibitemShut {NoStop}%
     \bibitem [{\citenamefont {${\rm \acute{A}}$lvarez}\ and\ \citenamefont {Alves}(2008)}]{lvarez_2008}%
       \BibitemOpen
       \bibfield  {author} {\bibinfo {author} {\bibfnamefont {R.}~\bibnamefont {${\rm \acute{A}}$lvarez}}\ and\ \bibinfo {author} {\bibfnamefont {L.~L.}\ \bibnamefont {Alves}},\ }\bibfield  {title} {\bibinfo {title} {Numerical solution to an electromagnetic model with neumann boundary conditions, for a microwave-driven plasma reactor},\ }\href {https://doi.org/10.1088/0022-3727/41/21/215204} {\bibfield  {journal} {\bibinfo  {journal} {Journal of Physics D: Applied Physics}\ }\textbf {\bibinfo {volume} {41}},\ \bibinfo {pages} {215204} (\bibinfo {year} {2008})}\BibitemShut {NoStop}%
     \bibitem [{\citenamefont {Vansteenkiste}\ \emph {et~al.}(2014)\citenamefont {Vansteenkiste}, \citenamefont {Leliaert}, \citenamefont {Dvornik}, \citenamefont {Helsen}, \citenamefont {Garcia~Sanchez},\ and\ \citenamefont {{Van Waeyenberge}}}]{Vansteenkiste2014}%
       \BibitemOpen
       \bibfield  {author} {\bibinfo {author} {\bibfnamefont {A.}~\bibnamefont {Vansteenkiste}}, \bibinfo {author} {\bibfnamefont {J.}~\bibnamefont {Leliaert}}, \bibinfo {author} {\bibfnamefont {M.}~\bibnamefont {Dvornik}}, \bibinfo {author} {\bibfnamefont {M.}~\bibnamefont {Helsen}}, \bibinfo {author} {\bibfnamefont {F.}~\bibnamefont {Garcia~Sanchez}},\ and\ \bibinfo {author} {\bibfnamefont {B.}~\bibnamefont {{Van Waeyenberge}}},\ }\bibfield  {title} {\bibinfo {title} {The design and verification of mumax3},\ }\href {https://doi.org/10.1063/1.4899186} {\bibfield  {journal} {\bibinfo  {journal} {AIP Advances}\ }\textbf {\bibinfo {volume} {4}},\ \bibinfo {pages} {107133} (\bibinfo {year} {2014})}\BibitemShut {NoStop}%
     \bibitem [{\citenamefont {Joos}\ \emph {et~al.}(2023)\citenamefont {Joos}, \citenamefont {Bassirian}, \citenamefont {Gypens}, \citenamefont {Mulkers}, \citenamefont {Litzius}, \citenamefont {Van~Waeyenberge},\ and\ \citenamefont {Leliaert}}]{joos_tutorial_2023}%
       \BibitemOpen
       \bibfield  {author} {\bibinfo {author} {\bibfnamefont {J.~J.}\ \bibnamefont {Joos}}, \bibinfo {author} {\bibfnamefont {P.}~\bibnamefont {Bassirian}}, \bibinfo {author} {\bibfnamefont {P.}~\bibnamefont {Gypens}}, \bibinfo {author} {\bibfnamefont {J.}~\bibnamefont {Mulkers}}, \bibinfo {author} {\bibfnamefont {K.}~\bibnamefont {Litzius}}, \bibinfo {author} {\bibfnamefont {B.}~\bibnamefont {Van~Waeyenberge}},\ and\ \bibinfo {author} {\bibfnamefont {J.}~\bibnamefont {Leliaert}},\ }\bibfield  {title} {\bibinfo {title} {Tutorial: {Simulating} modern magnetic material systems in mumax3},\ }\href {https://doi.org/10.1063/5.0160988} {\bibfield  {journal} {\bibinfo  {journal} {Journal of Applied Physics}\ }\textbf {\bibinfo {volume} {134}},\ \bibinfo {pages} {171101} (\bibinfo {year} {2023})}\BibitemShut {NoStop}%
     \bibitem [{\citenamefont {Xu}\ \emph {et~al.}(2023)\citenamefont {Xu}, \citenamefont {Li}, \citenamefont {Chen}, \citenamefont {Yu}, \citenamefont {Zhang}, \citenamefont {Wang},\ and\ \citenamefont {Wang}}]{fuming2023}%
       \BibitemOpen
       \bibfield  {author} {\bibinfo {author} {\bibfnamefont {F.~M.}\ \bibnamefont {Xu}}, \bibinfo {author} {\bibfnamefont {G.~Y.}\ \bibnamefont {Li}}, \bibinfo {author} {\bibfnamefont {J.}~\bibnamefont {Chen}}, \bibinfo {author} {\bibfnamefont {Z.~Z.}\ \bibnamefont {Yu}}, \bibinfo {author} {\bibfnamefont {L.}~\bibnamefont {Zhang}}, \bibinfo {author} {\bibfnamefont {B.~G.}\ \bibnamefont {Wang}},\ and\ \bibinfo {author} {\bibfnamefont {J.}~\bibnamefont {Wang}},\ }\bibfield  {title} {\bibinfo {title} {Unified framework of the microscopic {Landau-Lifshitz-Gilbert equation} and its application to skyrmion dynamics},\ }\href {https://doi.org/10.1103/PhysRevB.108.144409} {\bibfield  {journal} {\bibinfo  {journal} {Phys. Rev. B}\ }\textbf {\bibinfo {volume} {108}},\ \bibinfo {pages} {144409} (\bibinfo {year} {2023})}\BibitemShut {NoStop}%
     \bibitem [{\citenamefont {Lakshmanan}(2011)}]{lakshmanan_2011}%
       \BibitemOpen
       \bibfield  {author} {\bibinfo {author} {\bibfnamefont {M.}~\bibnamefont {Lakshmanan}},\ }\bibfield  {title} {\bibinfo {title} {The fascinating world of {Landau-Lifshitz-Gilbert Equation}: An overview},\ }\href {https://doi.org/10.1098/rsta.2010.0319} {\bibfield  {journal} {\bibinfo  {journal} {Philos. Trans. R. Soc. A: Math. Phys. Eng. Sci.}\ }\textbf {\bibinfo {volume} {369}},\ \bibinfo {pages} {1280} (\bibinfo {year} {2011})}\BibitemShut {NoStop}%
     \bibitem [{\citenamefont {Gilbert}(2004)}]{gilbert_2004}%
       \BibitemOpen
       \bibfield  {author} {\bibinfo {author} {\bibfnamefont {T.}~\bibnamefont {Gilbert}},\ }\bibfield  {title} {\bibinfo {title} {Classics in magnetics a phenomenological theory of damping in ferromagnetic materials},\ }\href {https://doi.org/10.1109/TMAG.2004.836740} {\bibfield  {journal} {\bibinfo  {journal} {{IEEE} Trans. Magn.}\ }\textbf {\bibinfo {volume} {40}},\ \bibinfo {pages} {3443} (\bibinfo {year} {2004})}\BibitemShut {NoStop}%
     \bibitem [{\citenamefont {Meisenheimer}\ \emph {et~al.}(2024)\citenamefont {Meisenheimer}, \citenamefont {Moore}, \citenamefont {Zhou}, \citenamefont {Zhang}, \citenamefont {Huang}, \citenamefont {Husain}, \citenamefont {Chen}, \citenamefont {Martin}, \citenamefont {Persson}, \citenamefont {Griffin}, \citenamefont {Caretta}, \citenamefont {Stevenson},\ and\ \citenamefont {Ramesh}}]{meisenheimer_2024}%
       \BibitemOpen
       \bibfield  {author} {\bibinfo {author} {\bibfnamefont {P.}~\bibnamefont {Meisenheimer}}, \bibinfo {author} {\bibfnamefont {G.}~\bibnamefont {Moore}}, \bibinfo {author} {\bibfnamefont {S.~Y.}\ \bibnamefont {Zhou}}, \bibinfo {author} {\bibfnamefont {H.~R.}\ \bibnamefont {Zhang}}, \bibinfo {author} {\bibfnamefont {X.~X.}\ \bibnamefont {Huang}}, \bibinfo {author} {\bibfnamefont {S.}~\bibnamefont {Husain}}, \bibinfo {author} {\bibfnamefont {X.~Z.}\ \bibnamefont {Chen}}, \bibinfo {author} {\bibfnamefont {L.~W.}\ \bibnamefont {Martin}}, \bibinfo {author} {\bibfnamefont {K.~A.}\ \bibnamefont {Persson}}, \bibinfo {author} {\bibfnamefont {S.}~\bibnamefont {Griffin}}, \bibinfo {author} {\bibfnamefont {L.}~\bibnamefont {Caretta}}, \bibinfo {author} {\bibfnamefont {P.}~\bibnamefont {Stevenson}},\ and\ \bibinfo {author} {\bibfnamefont {R.}~\bibnamefont {Ramesh}},\ }\bibfield  {title} {\bibinfo {title} {Switching the spin cycloid in {BiFeO$_3$} with an electric field},\ }\href {https://doi.org/10.1038/s41467-024-47232-5}
       {\bibfield  {journal} {\bibinfo  {journal} {Nat. Commun.}\ }\textbf {\bibinfo {volume} {15}},\ \bibinfo {pages} {2903} (\bibinfo {year} {2024})}\BibitemShut {NoStop}%
     \bibitem [{sup()}]{supply}%
       \BibitemOpen
       \href@noop {} {\bibinfo {title} {See the {Supplementary Material} online: {Video} 1 for the topological magnetic structures and the transition process as the magnetic field strength varies from 0 to -10 {T} in a $64\times64$ nm nanostructure; and {Video} 2 and 3 for the detailed switching processes between the trivial and the non-trivial topological configurations at the alternating magnetic fields with $f=1.5$ {GHz} and {$B_z$}=$-$4.1 {T}, and $f=3.0$ {GHz} and {$B_z$}=$-$4.5 {T}, respectively}}\BibitemShut {NoStop}%
     \bibitem [{\citenamefont {Dong}\ \emph {et~al.}(2021)\citenamefont {Dong}, \citenamefont {Tao}, \citenamefont {Duan}, \citenamefont {Li}, \citenamefont {Huang},\ and\ \citenamefont {Peeters}}]{Dong21}%
       \BibitemOpen
       \bibfield  {author} {\bibinfo {author} {\bibfnamefont {H.~M.}\ \bibnamefont {Dong}}, \bibinfo {author} {\bibfnamefont {Z.~H.}\ \bibnamefont {Tao}}, \bibinfo {author} {\bibfnamefont {Y.~F.}\ \bibnamefont {Duan}}, \bibinfo {author} {\bibfnamefont {L.~L.}\ \bibnamefont {Li}}, \bibinfo {author} {\bibfnamefont {F.}~\bibnamefont {Huang}},\ and\ \bibinfo {author} {\bibfnamefont {F.~M.}\ \bibnamefont {Peeters}},\ }\bibfield  {title} {\bibinfo {title} {Substrate dependent terahertz magneto-optical properties of monolayer ws2},\ }\href {https://doi.org/10.1364/OL.435055} {\bibfield  {journal} {\bibinfo  {journal} {Opt. Lett.}\ }\textbf {\bibinfo {volume} {46}},\ \bibinfo {pages} {4892} (\bibinfo {year} {2021})}\BibitemShut {NoStop}%
     \bibitem [{\citenamefont {Du}\ \emph {et~al.}(2015)\citenamefont {Du}, \citenamefont {Che}, \citenamefont {Kong}, \citenamefont {Zhao}, \citenamefont {Jin}, \citenamefont {Wang}, \citenamefont {Yang}, \citenamefont {Ning}, \citenamefont {Li}, \citenamefont {Jin}, \citenamefont {Chen}, \citenamefont {Zang}, \citenamefont {Zhang},\ and\ \citenamefont {Tian}}]{du_edge_2015}%
       \BibitemOpen
       \bibfield  {author} {\bibinfo {author} {\bibfnamefont {H.~F.}\ \bibnamefont {Du}}, \bibinfo {author} {\bibfnamefont {R.~C.}\ \bibnamefont {Che}}, \bibinfo {author} {\bibfnamefont {L.~Y.}\ \bibnamefont {Kong}}, \bibinfo {author} {\bibfnamefont {X.~B.}\ \bibnamefont {Zhao}}, \bibinfo {author} {\bibfnamefont {C.~M.}\ \bibnamefont {Jin}}, \bibinfo {author} {\bibfnamefont {C.}~\bibnamefont {Wang}}, \bibinfo {author} {\bibfnamefont {J.~Y.}\ \bibnamefont {Yang}}, \bibinfo {author} {\bibfnamefont {W.}~\bibnamefont {Ning}}, \bibinfo {author} {\bibfnamefont {R.~W.}\ \bibnamefont {Li}}, \bibinfo {author} {\bibfnamefont {C.~Q.}\ \bibnamefont {Jin}}, \bibinfo {author} {\bibfnamefont {X.~H.}\ \bibnamefont {Chen}}, \bibinfo {author} {\bibfnamefont {J.~D.}\ \bibnamefont {Zang}}, \bibinfo {author} {\bibfnamefont {Y.~H.}\ \bibnamefont {Zhang}},\ and\ \bibinfo {author} {\bibfnamefont {M.~L.}\ \bibnamefont {Tian}},\ }\bibfield  {title} {\bibinfo {title} {Edge-mediated skyrmion chain and its collective dynamics in a confined
       geometry},\ }\href {https://doi.org/10.1038/ncomms9504} {\bibfield  {journal} {\bibinfo  {journal} {Nat. Commun.}\ }\textbf {\bibinfo {volume} {6}},\ \bibinfo {pages} {8504} (\bibinfo {year} {2015})}\BibitemShut {NoStop}%
     \bibitem [{\citenamefont {Behera}\ \emph {et~al.}(2019)\citenamefont {Behera}, \citenamefont {Chowdhury},\ and\ \citenamefont {Das}}]{behera_magnetic_2019}%
       \BibitemOpen
       \bibfield  {author} {\bibinfo {author} {\bibfnamefont {A.~K.}\ \bibnamefont {Behera}}, \bibinfo {author} {\bibfnamefont {S.}~\bibnamefont {Chowdhury}},\ and\ \bibinfo {author} {\bibfnamefont {S.~R.}\ \bibnamefont {Das}},\ }\bibfield  {title} {\bibinfo {title} {Magnetic skyrmions in atomic thin {CrI}$_{\textrm{3}}$ monolayer},\ }\href {https://doi.org/10.1063/1.5096782} {\bibfield  {journal} {\bibinfo  {journal} {Appl. Phys. Lett.}\ }\textbf {\bibinfo {volume} {114}},\ \bibinfo {pages} {232402} (\bibinfo {year} {2019})}\BibitemShut {NoStop}%
     \bibitem [{\citenamefont {Yu}\ \emph {et~al.}(2018)\citenamefont {Yu}, \citenamefont {Morikawa}, \citenamefont {Yokouchi}, \citenamefont {Shibata}, \citenamefont {Kanazawa}, \citenamefont {Kagawa}, \citenamefont {Arima},\ and\ \citenamefont {Tokura}}]{yu_aggregation_2018}%
       \BibitemOpen
       \bibfield  {author} {\bibinfo {author} {\bibfnamefont {X.~Z.}\ \bibnamefont {Yu}}, \bibinfo {author} {\bibfnamefont {D.}~\bibnamefont {Morikawa}}, \bibinfo {author} {\bibfnamefont {T.}~\bibnamefont {Yokouchi}}, \bibinfo {author} {\bibfnamefont {K.}~\bibnamefont {Shibata}}, \bibinfo {author} {\bibfnamefont {N.}~\bibnamefont {Kanazawa}}, \bibinfo {author} {\bibfnamefont {F.}~\bibnamefont {Kagawa}}, \bibinfo {author} {\bibfnamefont {T.~H.}\ \bibnamefont {Arima}},\ and\ \bibinfo {author} {\bibfnamefont {Y.}~\bibnamefont {Tokura}},\ }\bibfield  {title} {\bibinfo {title} {Aggregation and collapse dynamics of skyrmions in a non-equilibrium state},\ }\href {https://doi.org/10.1038/s41567-018-0155-3} {\bibfield  {journal} {\bibinfo  {journal} {Nat. Phys.}\ }\textbf {\bibinfo {volume} {14}},\ \bibinfo {pages} {832} (\bibinfo {year} {2018})}\BibitemShut {NoStop}%
     \bibitem [{\citenamefont {Zheng}\ \emph {et~al.}(2023)\citenamefont {Zheng}, \citenamefont {Kiselev}, \citenamefont {Rybakov}, \citenamefont {Yang}, \citenamefont {Shi}, \citenamefont {Bl{\"u}gel},\ and\ \citenamefont {Dunin-Borkowski}}]{hopfion_2023}%
       \BibitemOpen
       \bibfield  {author} {\bibinfo {author} {\bibfnamefont {F.}~\bibnamefont {Zheng}}, \bibinfo {author} {\bibfnamefont {N.~S.}\ \bibnamefont {Kiselev}}, \bibinfo {author} {\bibfnamefont {F.~N.}\ \bibnamefont {Rybakov}}, \bibinfo {author} {\bibfnamefont {L.}~\bibnamefont {Yang}}, \bibinfo {author} {\bibfnamefont {W.}~\bibnamefont {Shi}}, \bibinfo {author} {\bibfnamefont {S.}~\bibnamefont {Bl{\"u}gel}},\ and\ \bibinfo {author} {\bibfnamefont {R.~E.}\ \bibnamefont {Dunin-Borkowski}},\ }\bibfield  {title} {\bibinfo {title} {Hopfion rings in a cubic chiral magnet},\ }\href {https://doi.org/10.1038/s41586-023-06658-5} {\bibfield  {journal} {\bibinfo  {journal} {Nature}\ }\textbf {\bibinfo {volume} {623}},\ \bibinfo {pages} {718} (\bibinfo {year} {2023})}\BibitemShut {NoStop}%
     \bibitem [{\citenamefont {Pathak}\ and\ \citenamefont {Hertel}(2021)}]{pathak_2021}%
       \BibitemOpen
       \bibfield  {author} {\bibinfo {author} {\bibfnamefont {S.~A.}\ \bibnamefont {Pathak}}\ and\ \bibinfo {author} {\bibfnamefont {R.}~\bibnamefont {Hertel}},\ }\bibfield  {title} {\bibinfo {title} {Three-dimensional chiral magnetization structures in {FeGe} nanospheres},\ }\href {https://doi.org/10.1103/PhysRevB.103.104414} {\bibfield  {journal} {\bibinfo  {journal} {Phys. Rev. B}\ }\textbf {\bibinfo {volume} {103}},\ \bibinfo {pages} {104414} (\bibinfo {year} {2021})}\BibitemShut {NoStop}%
     \bibitem [{\citenamefont {Qin}\ \emph {et~al.}(2022)\citenamefont {Qin}, \citenamefont {Zhang}, \citenamefont {Yang}, \citenamefont {Lv}, \citenamefont {Pei}, \citenamefont {Yang}, \citenamefont {Liu}, \citenamefont {Zhang},\ and\ \citenamefont {Che}}]{qin2c03046}%
       \BibitemOpen
       \bibfield  {author} {\bibinfo {author} {\bibfnamefont {G.}~\bibnamefont {Qin}}, \bibinfo {author} {\bibfnamefont {R.~X.}\ \bibnamefont {Zhang}}, \bibinfo {author} {\bibfnamefont {C.~D.}\ \bibnamefont {Yang}}, \bibinfo {author} {\bibfnamefont {X.~W.}\ \bibnamefont {Lv}}, \bibinfo {author} {\bibfnamefont {K.}~\bibnamefont {Pei}}, \bibinfo {author} {\bibfnamefont {L.~T.}\ \bibnamefont {Yang}}, \bibinfo {author} {\bibfnamefont {X.~H.}\ \bibnamefont {Liu}}, \bibinfo {author} {\bibfnamefont {X.~F.}\ \bibnamefont {Zhang}},\ and\ \bibinfo {author} {\bibfnamefont {R.~C.}\ \bibnamefont {Che}},\ }\bibfield  {title} {\bibinfo {title} {Magnetic-field-assisted diffusion motion of magnetic skyrmions},\ }\href {https://doi.org/10.1021/acsnano.2c03046} {\bibfield  {journal} {\bibinfo  {journal} {ACS Nano}\ }\textbf {\bibinfo {volume} {16}},\ \bibinfo {pages} {15927} (\bibinfo {year} {2022})}\BibitemShut {NoStop}%
     \end{thebibliography}
\end{document}